\documentclass[aps,twocolumn,nofootinbib,superscriptaddress,floatfix]{revtex4}
\usepackage{color,amsmath,amssymb,graphicx,latexsym,subfigure}
\usepackage{multirow}
\usepackage{amsmath}
\usepackage{bm}
\usepackage{float}
\usepackage{booktabs}
\usepackage{threeparttable}
\usepackage{CJK}
\usepackage[whole]{bxcjkjatype} 
\usepackage{CJKutf8}
\usepackage[colorlinks,linkcolor=green, anchorcolor=green,citecolor=green]{hyperref}
\usepackage{xcolor}

\usepackage{natbib}          

\renewcommand{\thetable}{\arabic{table}}  

\hypersetup{colorlinks=true,
	breaklinks=true,
	pdfstartview=Fit,
	linkcolor=blue,
	citecolor=blue,
	urlcolor=blue}

\begin{document}

\title[LPT]{A 44-minute periodic radio transient in a supernova remnant} 

\author{Di\,Li$^{\dagger}$}
\affiliation{Department of Astronomy, Tsinghua University, Beijing 100084, China.}
\affiliation{National Astronomical Observatories,Chinese Academy of Sciences,Beijing,100101,China.}
\affiliation{Zhejiang Lab, Hangzhou, Zhejiang 311121, China.}
\author{Mao\,Yuan$^{\dagger}$}
\affiliation{State Key Laboratory of Space Weather,Chinese Academy of Sciences, Beijing 100190, China.}
\affiliation{National Space Science Center, Chinese Academy of Sciences, Beijing 100190, China.}

\author{Lin\,Wu$^{\dagger}$}
\affiliation{State Key Laboratory of Space Weather,Chinese Academy of Sciences, Beijing 100190, China.}
\affiliation{National Space Science Center, Chinese Academy of Sciences, Beijing 100190, China.}

\author{Jing-ye\,Yan$^{\ast}$}
\affiliation{State Key Laboratory of Space Weather,Chinese Academy of Sciences, Beijing 100190, China.}
\affiliation{National Space Science Center, Chinese Academy of Sciences, Beijing 100190, China.}

\author{Xu-ning\,Lv$^{\ast}$}
\affiliation{State Key Laboratory of Space Weather,Chinese Academy of Sciences, Beijing 100190, China.}
\affiliation{National Space Science Center, Chinese Academy of Sciences, Beijing 100190, China.}

\author{Chao-Wei\, Tsai}
\affiliation{National Astronomical Observatories,Chinese Academy of Sciences,Beijing,100101,China.}
\affiliation{Institute for Frontiers in Astronomy and Astrophysics, Beijing Normal University, Beijing 102206, China.}

\author{Jia-qi\,Zhao}
\affiliation{Department of Astronomy, Tsinghua University, Beijing 100084, China.}

\author{Wei-wei\,Zhu}
\affiliation{National Astronomical Observatories,Chinese Academy of Sciences,Beijing,100101,China.}
\affiliation{Institute for Frontiers in Astronomy and Astrophysics, Beijing Normal University, Beijing 102206, China.}

\author{Li\,Deng}
\affiliation{State Key Laboratory of Space Weather,Chinese Academy of Sciences, Beijing 100190, China.}
\affiliation{National Space Science Center, Chinese Academy of Sciences, Beijing 100190, China.}

\author{Ai-lan\,Lan}
\affiliation{State Key Laboratory of Space Weather,Chinese Academy of Sciences, Beijing 100190, China.}
\affiliation{National Space Science Center, Chinese Academy of Sciences, Beijing 100190, China.}

\author{Ren-xin\,Xu}
\affiliation{Department of Astronomy, Peking University, Beijing 100871, China.}
\affiliation{Kavli Institute for Astronomy and Astrophysics, Peking University, Beijing 100871, China.}
\affiliation{State Key Laboratory of Nuclear Physics and Technology, School of Physics,Peking University,Beijing 100871, China.}

\author{Xiang-lei\,Chen}
\affiliation{National Astronomical Observatories,Chinese Academy of Sciences,Beijing,100101,China.}

\author{Ling-qi\,Meng}
\affiliation{National Astronomical Observatories,Chinese Academy of Sciences,Beijing,100101,China.}
\affiliation{School of Astronomy and Space Science, University of Chinese Academy of Sciences, Beijing 100049, China.}

\author{Jian\,Li}
\affiliation{Department of Astronomy, School of Physical Sciences, University of Science and Technology of China, Hefei 230026, China.}

\author{Xiang-dong\,Li}
\affiliation{School of Astronomy and Space Science, Nanjing University, Nanjing 210023, China.}
\affiliation{Key Laboratory of Modern Astronomy and Astrophysics (Nanjing University), Ministry of Education, Nanjing 210023, China.}

\author{Ping\,Zhou}
\affiliation{School of Astronomy and Space Science, Nanjing University, Nanjing 210023, China.}
\affiliation{Key Laboratory of Modern Astronomy and Astrophysics (Nanjing University), Ministry of Education, Nanjing 210023, China.}

\author{Hao-ran\,Yang}
\affiliation{School of Astronomy and Space Science, Nanjing University, Nanjing 210023, China.}
\affiliation{Key Laboratory of Modern Astronomy and Astrophysics (Nanjing University), Ministry of Education, Nanjing 210023, China.}

\author{Meng-yao\,Xue}
\affiliation{National Astronomical Observatories,Chinese Academy of Sciences,Beijing,100101,China.}

\author{Ji-guang\,Lu}
\affiliation{National Astronomical Observatories,Chinese Academy of Sciences,Beijing,100101,China.}

\author{Chen-chen\,Miao}
\affiliation{Zhejiang Lab, Hangzhou, Zhejiang 311121, China.}

\author{Wei-yang\,Wang}
\affiliation{Key Laboratory of Modern Astronomy and Astrophysics (Nanjing University), Ministry of Education, Nanjing 210023, China.}

\author{Jia-rui\,Niu}
\affiliation{National Astronomical Observatories,Chinese Academy of Sciences,Beijing,100101,China.}

\author{Zi-yao\,Fang}
\affiliation{National Astronomical Observatories,Chinese Academy of Sciences,Beijing,100101,China.}
\affiliation{School of Astronomy and Space Science, University of Chinese Academy of Sciences, Beijing 100049, China.}

\author{Qiu-yang\,Fu}
\affiliation{National Astronomical Observatories,Chinese Academy of Sciences,Beijing,100101,China.}
\affiliation{School of Astronomy and Space Science, University of Chinese Academy of Sciences, Beijing 100049, China.}

\author{Yi\, Feng}
\affiliation{Zhejiang Lab, Hangzhou, Zhejiang 311121, China.}

\author{Pei-jing\,Zhang}
\affiliation{Department of Physics, University of Helsinki, POBox 64 00014 Helsinki, Finland.}

\author{Jin-chen\,Jiang}
\affiliation{National Astronomical Observatories,Chinese Academy of Sciences,Beijing,100101,China.}

\author{Xue-li\,Miao}
\affiliation{National Astronomical Observatories,Chinese Academy of Sciences,Beijing,100101,China.}

\author{Yu\,Chen}
\affiliation{State Key Laboratory of Space Weather,Chinese Academy of Sciences, Beijing 100190, China.}
\affiliation{National Space Science Center, Chinese Academy of Sciences, Beijing 100190, China.}

\author{Ling-chen\,Sun}
\affiliation{State Key Laboratory of Space Weather,Chinese Academy of Sciences, Beijing 100190, China.}
\affiliation{National Space Science Center, Chinese Academy of Sciences, Beijing 100190, China.}
\author{Yang\,Yang}
\affiliation{State Key Laboratory of Space Weather,Chinese Academy of Sciences, Beijing 100190, China.}
\affiliation{National Space Science Center, Chinese Academy of Sciences, Beijing 100190, China.}
\author{Xiang\,Deng}
\affiliation{State Key Laboratory of Space Weather,Chinese Academy of Sciences, Beijing 100190, China.}
\affiliation{National Space Science Center, Chinese Academy of Sciences, Beijing 100190, China.}
\author{Shi\,Dai}
\affiliation{Western Sydney University, Locked Bag 1797, Penrith South DC, NSW 2751, Australia.}
\affiliation{School of Science, CSIRO Space and Astronomy, Australia Telescope National Facility, Epping, NSW 1710, Australia.}
\author{Xue\,Chen}
\affiliation{Beijing Institute of Technology Chongqing Innovation Center, Chongqing, China.}
\author{Ju-mei\,Yao}
\affiliation{Xinjiang Astronomical Observatory, Chinese Academy of Sciences, Urumqi, Xinjiang 830011, China.}
\author{Yu-jie\,Liu}
\affiliation{State Key Laboratory of Space Weather,Chinese Academy of Sciences, Beijing 100190, China.}
\affiliation{National Space Science Center, Chinese Academy of Sciences, Beijing 100190, China.}
\author{Chang-heng\,Li}
\affiliation{State Key Laboratory of Space Weather,Chinese Academy of Sciences, Beijing 100190, China.}
\affiliation{National Space Science Center, Chinese Academy of Sciences, Beijing 100190, China.}
\author{Ming-lu\,Zhang}
\affiliation{State Key Laboratory of Space Weather,Chinese Academy of Sciences, Beijing 100190, China.}
\affiliation{National Space Science Center, Chinese Academy of Sciences, Beijing 100190, China.}
\author{Yi-wen\,Yang}
\affiliation{State Key Laboratory of Space Weather,Chinese Academy of Sciences, Beijing 100190, China.}
\affiliation{National Space Science Center, Chinese Academy of Sciences, Beijing 100190, China.}
\author{Yu-cheng\,Zhou}
\affiliation{State Key Laboratory of Space Weather,Chinese Academy of Sciences, Beijing 100190, China.}
\affiliation{National Space Science Center, Chinese Academy of Sciences, Beijing 100190, China.}
\author{Yi-yi\,zhou}
\affiliation{State Key Laboratory of Space Weather,Chinese Academy of Sciences, Beijing 100190, China.}
\affiliation{National Space Science Center, Chinese Academy of Sciences, Beijing 100190, China.}
\author{Yong-kun\,Zhang}
\affiliation{National Astronomical Observatories,Chinese Academy of Sciences,Beijing,100101,China.}
\author{Chen-hui\,Niu}
\affiliation{Center China Normal University, Wuhan 430079, China.}
\author{Ru-shuang\,Zhao}
\affiliation{Guizhou Normal University, Guiyang 550001, China.}
\author{Lei\,Zhang}
\affiliation{National Astronomical Observatories,Chinese Academy of Sciences,Beijing,100101,China.}
\affiliation{Centre for Astrophysics and Supercomputing, Swinburne University of Technology, P.O. Box 218, Hawthorn, VIC 3122, Australia.}
\author{Bo\,Peng}
\affiliation{National Astronomical Observatories,Chinese Academy of Sciences,Beijing,100101,China.}
\author{Ji\,Wu}
\affiliation{State Key Laboratory of Space Weather,Chinese Academy of Sciences, Beijing 100190, China.}
\affiliation{National Space Science Center, Chinese Academy of Sciences, Beijing 100190, China.}
\author{Chi\,Wang}
\affiliation{State Key Laboratory of Space Weather,Chinese Academy of Sciences, Beijing 100190, China.}
\affiliation{National Space Science Center, Chinese Academy of Sciences, Beijing 100190, China.}
\begin{abstract}

Long-period radio transients (LPTs) are a newly discovered class of radio emitters with periods ranging from minutes to hours. The astrophysical nature remains undetermined, particularly of LPTs with no detectable companions. We report the first evidence for a plausible supernova remnant (SNR) association with an LPT  (DART J1832-0911, 2656.23±0.15~s period), which supports a neutron star origin of such objects. The dispersion measure of this LPT, SNR's CO emission and HI absorption, and low probability of chance of alignment with field pulsars are all consistent with such an association.  The source displays either phase-locked circular or nearly 100\% linear polarization, indicating its strong and geometrically stable magnetic field. No detectable optical counterpart was found, even with a 10m-class telescope. The SNR association and the stable polarization suggest that DART J1832-0911 most likely originates from a young neutron star, whose spin could have been braked by supernova’s fallback materials. This discovery provides critical insights into the nature of ultra-long period transients and their link to stellar remnants.

\end{abstract}

\begin{center}
\begin{minipage}{0.9\linewidth}
$\dagger$ {These authors contributed equally to this work}\\
* Corresponding authors: yanjingye@nssc.ac.cn
\end{minipage}
\end{center}

\maketitle

\section{Introduction}\label{intro.sec}

Long-period radio transients (LPTs) are a new class of periodic radio emitters, characterized by ultra-long rotation periods and strong magnetic fields. These features place LPTs significantly outside the conventional framework of rotation-powered neutron star pulsars or white dwarfs \citep{marsh2016radio,pelisoli20235,hurley2022radio,hurley2023long,caleb2024emission,rea2024long,de2025sporadic,lee2025emission,dong2025chime,bloot2025strongly}. Among reported LPT cases, 
several are identified, or argued to originate from,  white-dwarf binary systems \cite{marsh2016radio,pelisoli20235,de2025sporadic,hurley20242,dong2025chime,bloot2025strongly}, including both spin-down powered white-dwarf pulsars (i.e. AR Sco) and accretion-powered systems. Another group shows no compelling evidence for companions: GLEAM-X J162759.5-523504.3 (18.18 min) \cite{hurley2022radio}, GPM J1839-10 (21 min) \cite{hurley2023long}, ASKAP J1935+2148 (53 min) \cite{caleb2024emission}, CHIME J0630+25 (7~min) \cite{dong2024discovery}. For the stellar candidates LPTs without confirmed companions,  the two primary models---spin-down powered pulsars and white dwarfs---each face significant challenges in explaining the observed emissions if they independently evolve in a traditional way: losing rotation energy through radio emission. Neutron stars with ultra-long periods typically fall below the so-called ``death line'', where their rotational energy becomes insufficient to sustain coherent radio emission \cite{chen1993pulsar,young1999radio,caleb2022discovery}, while white dwarfs often have insufficient magnetic fields to produce the observed periodic, coherent emissions \cite{rea2024long}. Although theoretical models lean toward isolated LPTs being magnetars (i.e. neutron stars with strong magnetic field) \cite{cooper2024beyond,men2025highly,beniamini2023evidence}, direct observational evidence and evolution studies remain limited.

\section{Observation and Results}

\subsection{DART Observation}
We report a new LPT in a supernova remnant, DART J1832-0911, detected by DAocheng Radio Telescope (DART) (see supplementary material)  with interferometric imaging across a frequency range of 149-459 MHz. The first pulse was detected on Modified Julian Date (MJD) 60374. From MJD 60401 to MJD 60443, we conducted continuous observations of the transient using an interferometric imaging technique with full polarization, simultaneously recording data at two distinct center frequencies, each with a bandwidth of 4~MHz. The recorded pulses showed an estimated peak flux density between  0.5--2~Jy. After this period, the transient entered a long-period quiescent state, as recorded by DART observation. 

\begin{figure}
  \centering
  \includegraphics[width=0.98\linewidth]{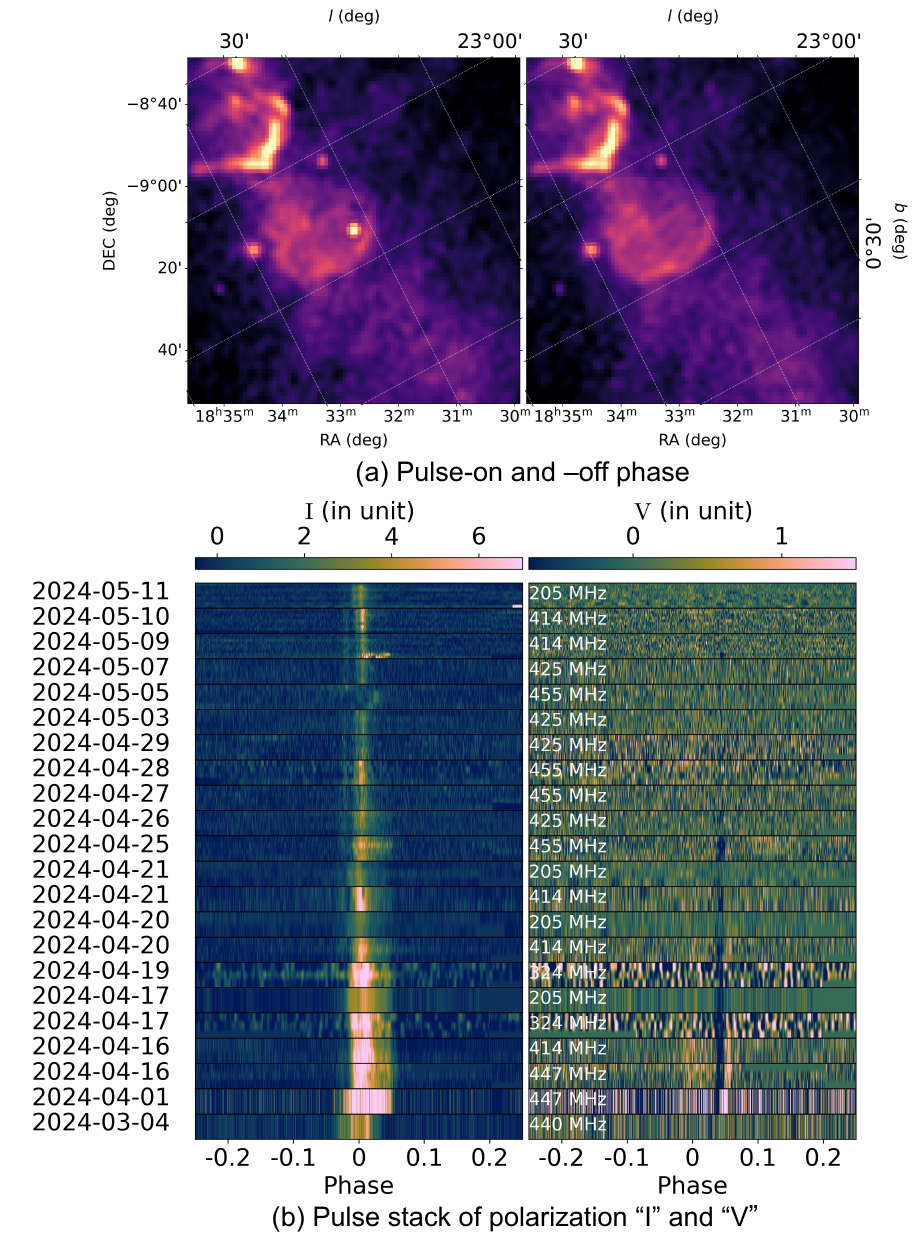}
  \caption{\textbf{ DART observation of J1832-0911.} (a)The top panel displays the radio image of DART J1832-0911 during its pulse-on (150~s duration) and pulse-off (375~s duration) phases, observed on 2024-04-16 at 425 MHz. The source is centered within the supernova remnant SNR G22.7-0.2.  The bottom panel shows the pulse stack of each observation epoch, with Stokes~I (left) and Stokes~V (right). Each row is an intensity map (phase vs.\ pulse number) from a single observing session, arranged in chronological order by MJD (labeled on the left). The observed frequency band for each session is labeled on the right.}
\label{supernova}
\end{figure}

\begin{figure}
  \centering
  \includegraphics[width=0.98\linewidth]{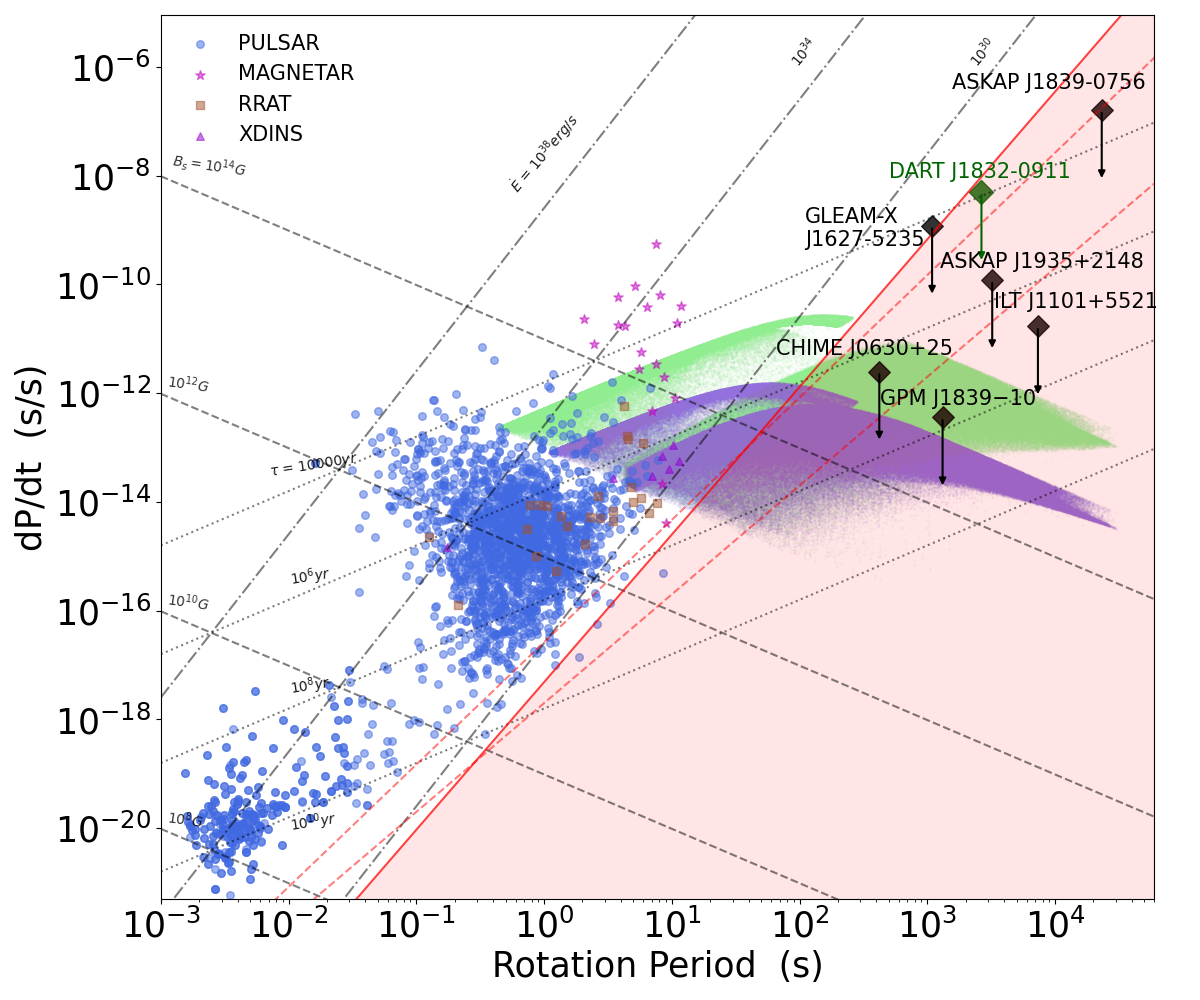}
\caption{\textbf{P-Pdot diagram of radio transients under a spin-down powered scenario.} DART J1832-0911 and other reported isolated long-period radio transients are marked by diamonds with arrows indicating their upper limits on period derivatives. The other types of transients are pulsars (blue dots), magnetars (purple stars), RRATs (brown squares), and X-ray Dim Isolated Neutron Stars (XDINS, light purple triangles). The solid red line represents the theoretical death line for spin-down powered neutron stars with a pure dipole magnetic field \cite{chen1993pulsar}, while the dashed red lines show the death lines for neutron stars with multipole fields \cite{zhang2000radio}. The green and purple shaded areas indicate the period ranges influenced by supernova fallback disk accretion from simulations (see supplementary material). These regions correspond to supernova remnant ages of $3\times 10^{4}$ and $2\times 10^{5}$~years. }
 \label{ppdot}
\end{figure}

The interferometric image allowed us to determine the celestial coordinates to be RA=18h32m48.5s ($\pm{5^{''}}$) and DEC=-09d11m17s ($\pm{5^{''}}$), situating it within the projected region of supernova remnant G22.7-0.2. Top panel in Fig.\ref{supernova} displays the pulse-on image of DART J1832-0911 at 425 MHz. The bottom panel shows the observed single pulses in two Stokes components. Additional pulse profile shapes are provided in the supplementary material (Fig.\ref{profile}). All pulse series presented in this study have been aligned through periodic folding and corrected for delay time. By analyzing the time of arrival (TOA) of single pulses recorded by DART, we derive a spin period of $P = 2656.23\pm0.15$~s with an estimated period derivative of $\dot{P}$ $< 5\times 10^{-9}$~ s/s. Fig.\ref{ppdot} displays the $P$-$\dot{P}$ diagram of DART J1832-0911 as well as other LPTs.

\subsection{FAST Observation}
To further investigate, we used the Five-hundred-meter Aperture Spherical Telescope (FAST) to observe the source at L-band during DART's non-detection period on MJD 60559. During a 1-hour observation, we detected a short pulse with a peak flux density of approximately 94~mJy. The duration of the pulse (i.e. the pulse width) is 0.2~s, just 0.007\% of the rotation period. This could point to a highly energetic and sharp emission event occurring at the strong magnetic field, like the dwarf pulse emission process of a neutron star \cite{chen2023strong,yan2024dwarf}.
\begin{figure}
  \centering
  \includegraphics[width=0.98\linewidth]{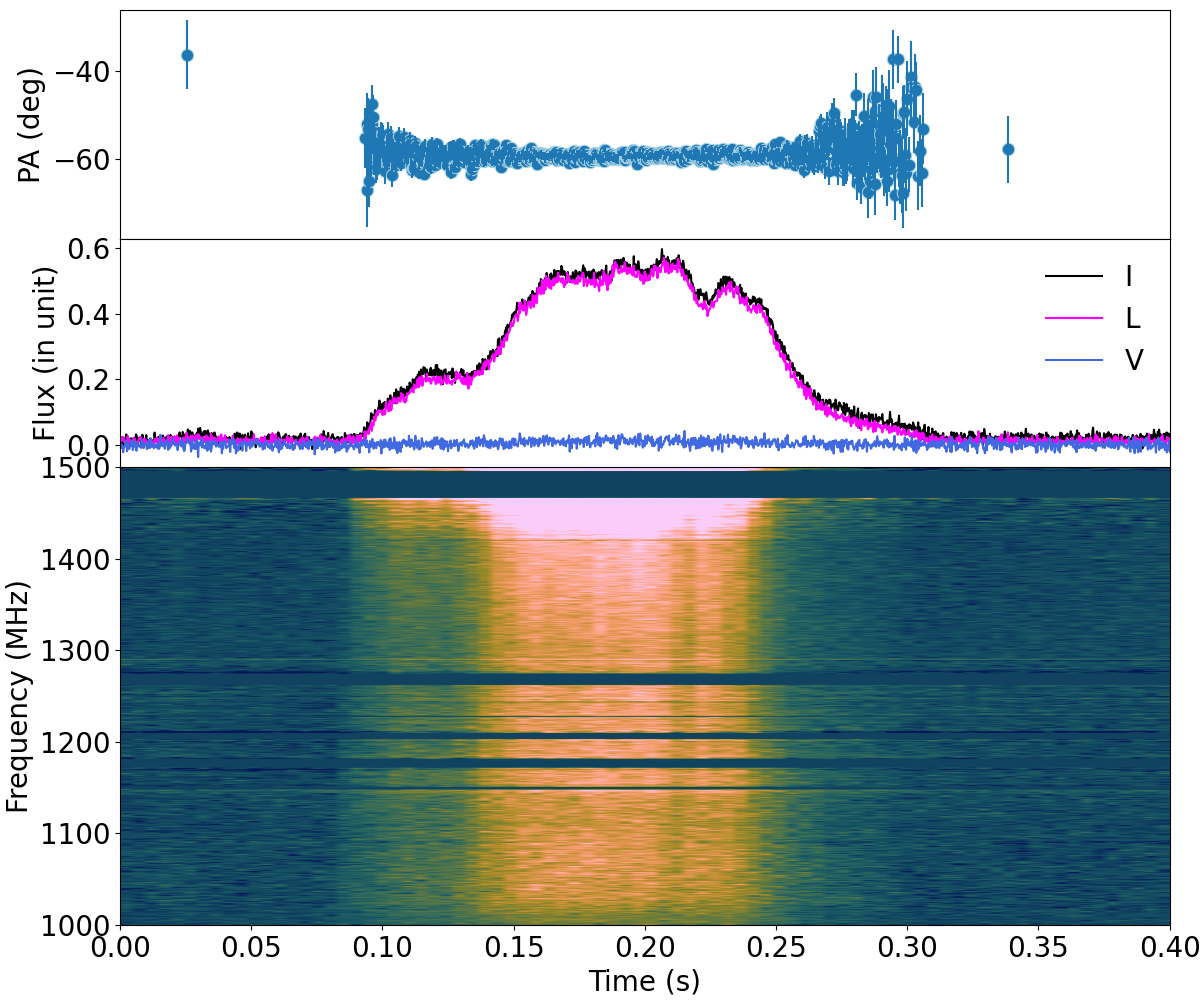}
 \caption{\textbf{The short pulse detected by FAST.} The top panel displays the linear polarization angle, while the middle panel presents the pulse profiles, including the total intensity(I), linear polarization(L), and circular polarization(V).}
 \label{fast}
\end{figure}

We folded the time series based on the ephemeris derived from DART observations and found that the TOA of this pulse is within the predicted emission phase. This short pulse is shown in  Fig.\ref{fast}.  The rotation measure (RM) of this pulse is $105.6 \pm 0.1\textrm{rad m}^{-2}$. A summary of the measured parameters of this transient is provided in Table.\ref{tab:para}.
\begin{table*} 
	\centering
	\caption{\textbf{Measured parameters of DART J1832-0911 from the DART and FAST observation.}}
	\label{tab:para} 
	
	\begin{tabular}{lcc} 
		\\
		\hline
		Parameter & DART& FAST\\
		          &149-459~MHz &1-1.5~GHz\\
		\hline
		Period derivative (s/s)& $<5\times 10^{-9}$&-\\
		DM ($\textrm{pc cm}^{-3}$) & 480$\pm{136}$ & 464.5$\pm{0.7}$\\
		RM ($\textrm{rad m}^{-2}$) &-&$+105.6\pm{0.1}$\\
		Pulse Width (s)  &40-250&0.2\\
		Peak flux density (Jy)  &0.5-2 &0.096$\pm$0.027\\
		\hline
	\end{tabular}
\end{table*}
\subsection{Optical and X-ray Counterpart Search}
We conducted an optical observation using the Gran Telescopio Canarias (GTC) on 30 July 2024. The integration time for this observation was 1.25 hours. 

\begin{figure}
  \centering
\includegraphics[width=0.98\linewidth]{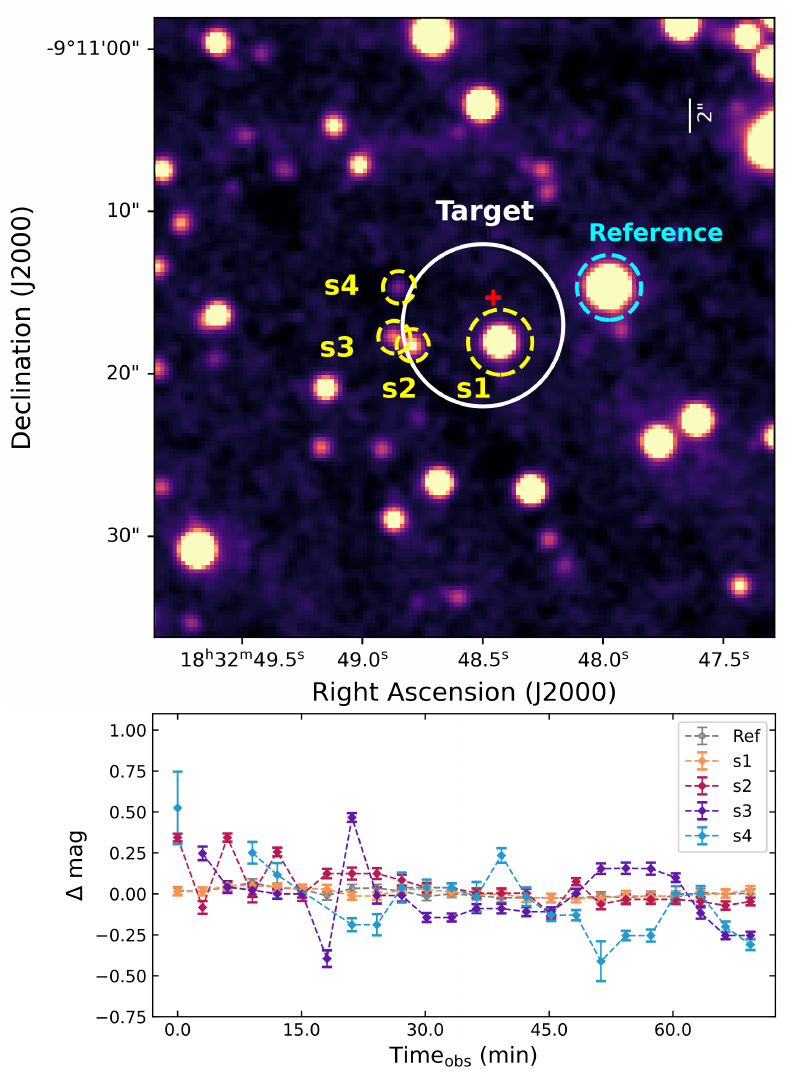}
\caption{\textbf{Optical search of the DART source.} The top panel shows the GTC $r$-band image of the source's celestial field. The white circle indicates the positional uncertainty of DART J1832-0911, with a range at $5^{\prime \prime}$.  The red cross in the circle indicate the accurate location, observed by MeerKAT \cite{wang2025detection}. The bottom panel presents the optical light curves of the reference star and a nearby candidate marked in the top panel. Neither light curve exhibits periodicity or variation timescale comparable to that of DART J1832-0911.}
 \label{opt}
\end{figure}

No credible optical counterpart was identified in the GTC observation, effectively ruling out any pulsating white dwarf with a temperature $<$~17,000K as its origin. An \textit{r}-band image of the DART source location was obtained using the broad-band imaging mode of OSIRIS$+$ on the GTC. As shown in Fig.\ref{opt}, there are four optical candidates within the designated zone. However, the light curves of these candidates show no periodic variations on timescales of tens of minutes, indicating they are not associated with DART J1832-0911. Given an upper limit of 26.6 magnitudes for the GTC observations and assuming a white dwarf radius of $r=0.2 \rm R_{\odot}$ at a distance of 4.5~kpc, any WD detectable by the GTC would have an effective temperature below approximately $\rm T_{eff} \sim 17000 K$, indicating an evolution age of hundreds of millions years (see supplementary material). The absence of an optical counterpart supports the possibility of a compact neutron star rather than a WD. Additionally, our analysis shows that a spin-down powered WD would be unable to account for the observed emission power of a transient with characteristics like those of DART J1832-0911  (see supplementary material).
We searched for any possible X-ray counterpart using archival data from the XMM-Newton observatory, obtained on March 13, 2011 (Observation ID: 0654480101). The observation had an exposure time of $1.54\times10^{4}$~s and covered the source with both XMM-PN and XMM-MOS2 detectors in the 0.2-10~keV. Within the positional uncertainty of DART J1832-0911, only a single photon was detected by XMM-PN. Consequently, no X-ray counterpart was identified for either the transient or the associated SNR region according to the observation in 2011. This yields an upper limit of $6.648\times10^{-14}$~erg~cm$^{-2}$~s$^{-1}$ in 0.2-10 keV at a 99\% confidence level. 

Independently of our work, a recent study \cite{wang2025detection}  has reported a 44-min periodic radio transient in the same direction, consistent with our detection (referred to as ASKAP~J1832$-$0911 in their work). The authors performed a comprehensive X-ray analysis of J1832$-$0911 using additional instruments, including {\it Chandra}, {\it Swift}, and Einstein Probe \cite{Yuan2015EP}, leading to the identification of a transient X-ray counterpart detected by {\it Chandra}. Their observations revealed an extreme contrast between a bright X-ray outburst ($L_{\mathrm{X}} \approx 7\times10^{32}$ erg s$^{-1}$) and a deep quiescent state ($L_{\mathrm{X}} < 10^{32}$ erg s$^{-1}$). Assuming a distance of 4.5 kpc, these correspond to flux limits of approximately $2–3\times10^{-14}$ erg cm$^{-2}$ s$^{-1}$ in the 1–10~keV band. We suggest that rapid follow-up X-ray observations conducted during the radio-bright phases of future detected LPTs will be critical for unveiling their physical nature.

Archival XMM-Newton observations from 2011 revealed no detectable X-ray emission from DART J1832–0911, whereas an X-ray burst was detected by Chandra in February 2024 but not in August \citep{wang2025detection}, indicating that J1832–0911 is an intermittent X-ray transient.  

\subsection{Dispersion Measure and SNR Association Analysis}
We calculated the dispersion measure (DM) using dual-frequency pulse series from DART and the short pulse spectra of FAST. By relating the TOA delay to the dispersion effect observed at two frequencies simultaneously recorded, we obtained a DM of $480\pm136$ $\textrm{pc}$ $\cdot\textrm{cm}^{-3}$. A DM search of the FAST spectra yielded $464.5\pm0.7$ $\textrm{pc} \cdot\textrm{cm}^{-3}$. These measurements from the two observatories are consistent within the 3~sigma error. Given the broader bandwidth and finer temporal resolution of FAST, we consider this result to be the precise DM for DART J1832-0911. 

We estimated the DM distance of DART J1832-0911, it is consistent with the distance of SNR G22.7-0.2. Previous studies estimate the distance to SNR G22.7-0.2 at 4.4$\pm 0.4$~kpc \cite{su2014interaction} or 4.7$\pm 0.2$~kpc\cite{ranasinghe2018revised}, based on molecular cloud analyses. Meanwhile, the DM-derived distance to DART J1832-0911 was estimated using three Galactic free-electron density models: YMW16 \cite{yao2017new}, NE2001 \cite{cordes2002ne2001}, and NE2025 \cite{ocker2026ne2025}, yielding distances of approximately 4.5 kpc, 5.9 kpc, and 6.4 kpc, respectively. All three distances are broadly consistent with the SNR distance given the known systematic uncertainties of free-electron density models, which can reach 30–50\%. This suggests that J1832-0911 may reside within the SNR bubble.  
Nevertheless, the DM-estimated distance remains debated due to the electron-density model dependence. 

To evaluate the reliability of DM-based distance estimation along this line of sight, we compare DM-derived distances with VLBI parallax distances with two pulsars \cite{deller2019microarcsecond}，which are located within $\pm5^\circ$ in declination of DART J1832$-$0911: PSR~J1820$-$0427 and PSR~J1901$-$0906. For PSR~J1820$-$0427 (DM $= 84.4$~pc~cm$^{-3}$), the NE2001, YMW16, and NE2025 models yield distances of 1.94, 2.92, and 2.01~kpc, respectively, while a VLBI parallax distance is $2.85^{+0.52}_{-0.35}$~kpc. For PSR~J1901$-$0906 (DM $= 72.7$~pc~cm$^{-3}$), the DM models give 2.13, 2.89, and 2.43~kpc, and the VLBI distance is $1.96^{+0.17}_{-0.23}$~kpc. 
We estimate the uncertainty in DM-derived distances based on the VLBI measurements using two approaches. First, we evaluate each of the three DM models individually.  The average fractional deviations between the DM distances and the VLBI distances for the two sightlines are $\sim$20\% for NE2001, $\sim$25\% for YMW16, and $\sim$27\% for NE2025. Based on these model-specific uncertainties, the DM-derived distances of DART~J1832$-$0911 are 4.7--7.1~kpc (NE2001), 3.4--5.6~kpc (YMW16), and 4.7--8.1~kpc (NE2025). Taking the full range of these estimates, we adopt a DM distance of 3.4 to 8.1~kpc. The second estimation method combines all three DM models. We take the 
mean of the three model predictions as the reference distance, $\bar{d} = (4.5 + 5.9 + 6.4)/3 \approx 5.6$~kpc. The total uncertainty is estimated by 
combining in quadrature the mean single-model systematic 
uncertainty ($\sim$24\%) and the fractional model-to-model dispersion ($\sigma_{\rm model} = (6.4-4.5)/5.6 \approx 34\%$), yielding a combined fractional uncertainty of 
$\sqrt{24^2 + 34^2} \approx 42\%$, or $\sim$2.4~kpc. This gives a consolidated distance range for DART~J1832$-$0911 of 3.2--8.0~kpc. The independently determined distance to SNR~G22.7$-$0.2 of 4.4--4.7~kpc \cite{su2014interaction,ranasinghe2018revised} falls within both the individual-model ranges and the consolidated interval, supporting a physical association between DART~J1832$-$0911 and the SNR.  
 
For J1832–0911, a VLBI observation can provide a geometric, model-independent method for distance measurement. In principle, this technique could be effective for bright and periodically active LPTs. However, several practical challenges remain. Parallax measurements require multiple epochs distributed over at least one year to disentangle parallax and proper motion\cite{chatterjee2009precision,deller2019microarcsecond}, which is difficult for J1832-0911 given its long-term nulling and variable emission strength. Although single-epoch localization is feasible, consistent multi-epoch detections are necessary for reliable distance determination. In addition, an excessively long or short emission window can reduce phase stability across long baselines\cite{chatterjee2004pulsar}. Therefore, while VLBI remains the most direct and conclusive approach for testing an LPT-SNR association, its application is presently constrained by observational scheduling and the source’s intermittent visibility.
 
We inspected the CO observations toward SNR G22.7-0.2. The velocity of CO emission ranges from 75 to 79 km/s \cite{su2014interaction}, consistent with the $\mathrm{HI}$ absorption ($75\pm15$ km/s) reported in Wang et al. \cite{wang2025detection}. Such alignment in the gas dynamics strongly supports that the compact object and the SNR reside within the same large-scale ISM structure.

\begin{figure*}
  \centering
  \includegraphics[width=0.6\linewidth]{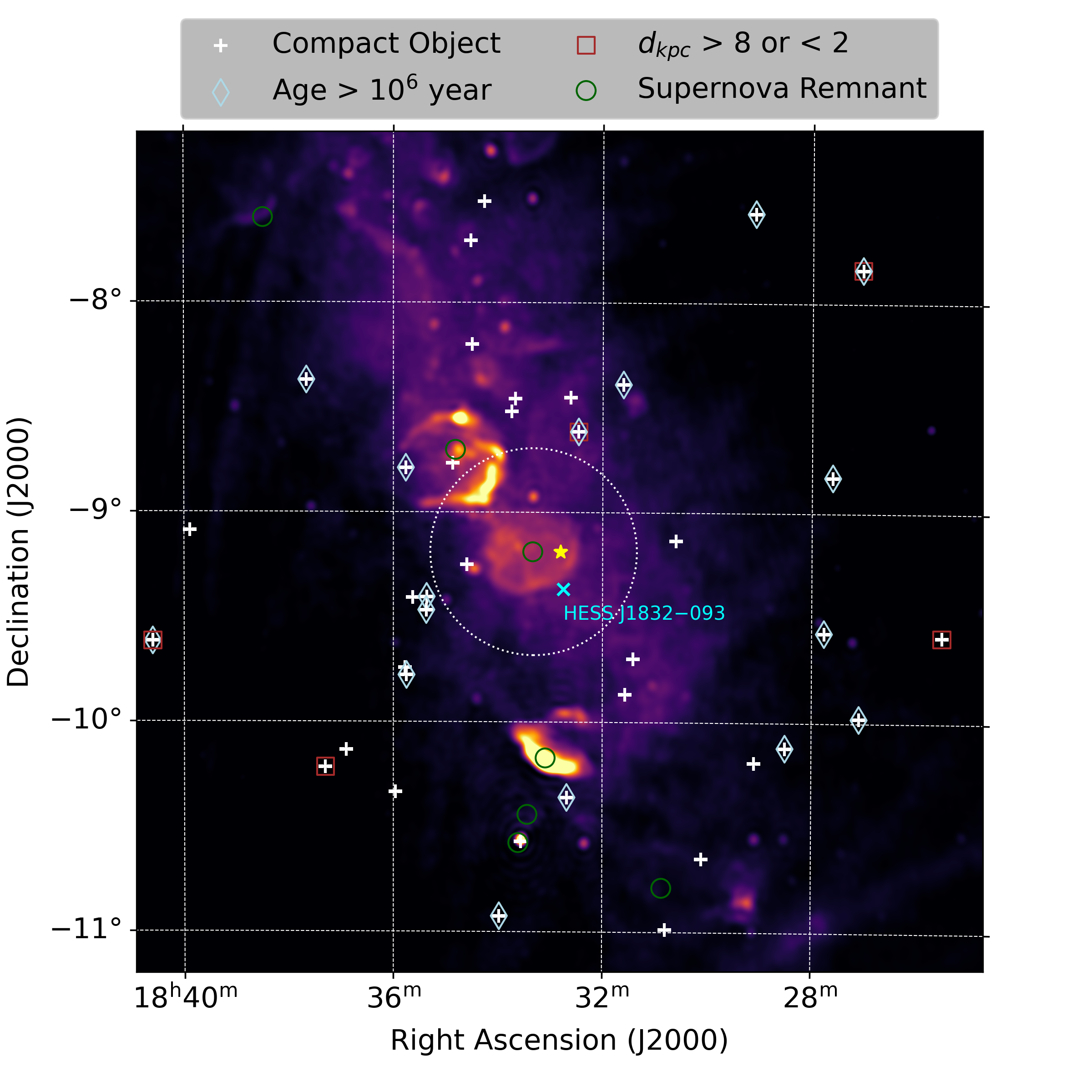}
 \caption{\textbf{The SNRs and compact objects around DART J1832-0911, illustrated by the yellow star.} The white dotted circle defines a reasonable limits of a projected escaping plane for a neutron star born in the center of G22.7-0.2. The radius of this region is 40~pc.}
 \label{snr_psr}
\end{figure*}

We inspect all compact objects within the 16 deg$^2$ region FoV of our DART observation. There are 17 other sources, which fall within the DM distance of 2 kpc to 8 kpc and an age below 1 million years, accommodating a ~100\% uncertainty allowed for the DM uncertainty and the SNR distance, and a 1000\% uncertainty allowed for the upper limits of the SNR age, as indicated in the Fig.\ref{snr_psr}. The random chance of any of these sources to get within the distance between the LPT and the SNR center is lower than 0.7\% (see supplementary material). Note that these sources have a chance of being SNR related. The contamination from the field pulsar population has to be much lower. And the previous pulsar (PSR) -SNR association studies reported no case of pulsars discovered around SNR to be later revealed to a chance coincidence (see supplementary material for a detail discussion).

We also investigated the X-ray emission of SNR G22.7-0.2 using data from XMM-Newton and previous studies. Our analysis, consistent with earlier research, reveals no evidence of extended X-ray emission within the radio range . The cessation of X-ray emission from the SNR suggests that the remnant's shock velocity, $V_{s}$, has decreased below 200~km/s---insufficient to heat the gas to X-ray-emitting temperatures \cite{mckee1974x,chevalier1999supernova}. Consequently, we can reasonably infer that the SNR has entered the radiative phase, during which the age of the SNR can be estimated as $t\sim 0.31 R/V > 3\times10^{4}$ years where $R=18$~pc is the SNR radius. Assuming a lower limit for the shock velocity of 30 km/s which is the minimum shock velocity required to produce the observed high-velocity wings and redshifted broadening in the CO emission of this SNR \cite{su2014interaction}, we obtain an upper age limit of $t<2\times10^{5}$ years.
 Assuming that DART~J1832$-$0911 is associated with SNR~G22.7$-$0.2, we use the age and radius of the SNR 
to estimate its projected transverse velocity to be 
26--176~km/s (see Section~S6 for details), which aligns well with the velocity range of the SNR-associated pulsars\cite{malov2021pulsars}.

Conclusively, the widely used association judgement: positional overlap,  distance, proper motion, and the 
consistency between the SNR age and the spin-down timescale 
expected from fallback disk models, shows no evidence against the physical association between DART J1832-0911 and SNR G22.7-0.2. Nevertheless, further strong evidence from direct observational measurements is required .  

\subsection{Radio Emission Properties}
DART J1832-0911 displays a range of emission characteristics during its active radio period. Prior to either undetectable weak emission or nulling, the source undergoes mode changes modulated by variations in pulse width and strength, as illustrated in the bottom panel of Fig.\ref{supernova}. 
These time-dependent mode changes reveal an evolution from bright, wide pulses to weaker, narrower ones. In the wide mode, pulse widths range from  100~s to 300~s, with a flux density of about 2~Jy. In the narrow mode, pulse width narrows to approximately 40~s, and flux density  decreases to roughly 0.5~Jy. During the emission-quiescent state, as observed by DART, we captured a short-lived pulse by FAST. Within the expected periodic phase, this pulse lasted 0.2~s and exhibited a peak flux of $\sim 94\pm 27$~mJy. Such brief, isolated emission components have also been observed in other long-period transients \cite{hurley2023long,caleb2024emission}. 

The polarization recorded of DART J1832-0911 presents unique properties as it significantly diverges from the other LPTs. While all reported isolated LPTs---except CHIME J0630+25 for which the research provided no polarization information---exhibit notable polarized components throughout their emission cycles, DART's pulses show a sporadic distribution of polarization components. Additionally, many pulses are extremely poorly polarized throughout the entire phase. Aside from a few sharp impulsive linear polarization (LP) components, we found no other LP emissions with a signal-to-noise ratio (S/N) greater than $3\sigma$ for most pulses.  Fig.\ref{lp-burst} highlights two instances of narrow LP components with S/N of 14.2~$\sigma$ (MJD 60416) and 6.6~$\sigma$ (MJD 60421), respectively. Regarding circular polarization (CP), we observed that pulses in the wide mode exhibit narrow and phase-locked CP components, as illustrated in the bottom panel of Fig.\ref{supernova}. The peak flux ratio of the highest CP components to the aligned total intensity reaches V/I$\sim 64\%$. Additionally, most CP components display sign reversal (see  Fig.\ref{lp-burst}). The narrow width of the CP components, spanning less than 3\% of the period, emphasizes a fixed circularly polarized emission zone in a limited region. 
Notably, the short-lived pulse detected by FAST is nearly 100\% linearly polarized, and the PA is flat across the whole emission duration. Fig.\ref{fast} shows the dynamic spectra and polarization properties of this pulse.

\section{Conclusion}
We report the discovery of a long-period radio transient, DART~J1832$-$0911, detected with interferometric imaging by the Daocheng Radio Telescope (DART) across 149--459~MHz. The spin period of this LPT is $P=2656.23\pm0.15$~s (44.27~min).
The interferometric localization (RA $=18^{\rm h}32^{\rm m}48.5^{\rm s}$, Dec $=-09^\circ11'17''$, $\pm5''$) situates DART~J1832$-$0911 within the projected extent of the supernova remnant G22.7$-$0.2. 
During the DART non-detection epoch, a short pulse detected by FAST at L-band occurred at the predicted rotational phase and is nearly $100\%$ linearly polarized with a well-measured RM, demonstrating that the source can emit extremely narrow, highly polarized bursts despite long intervals of radio quiescence.

The DM measurements of J1832-0911 from DART and FAST are roughly consistent, and  the DM-derived distance ($\sim4.5$,$\sim5.9$ and $\sim6.4$~kpc from YMW16, NE2001, and NE2025 models, respectively) are broadly consistent with independent distance estimates for SNR~G22.7$-$0.2, supporting a physical association between the compact object and the remnant environment. Complementary CO kinematics and H{\sc I} absorption arguments further strengthen the case that both the SNR and the transient reside in the same large-scale ISM structure. 
A field-population analysis yields a low probability for a chance alignment with unrelated compact objects, adding statistical weight to the association scenario. 

DART~J1832$-$0911 shows distinctive radio emission and polarization phenomenology. In DART bands, the activity evolves between wider and narrower emission modes and ultimately into a long quiescent state, with peak flux densities of order 0.5--2~Jy during active epochs. Polarization properties are notably intermittent: many pulses are weakly polarized, while (i) narrow, phase-locked circular polarization components appear in the wide mode with $V/I$ reaching $\sim64\%$ and occasional sign reversals, indicating a stable and geometrically confined emission region; and (ii) the FAST burst shows nearly pure linear polarization with a flat position angle across the burst window, pointing to an ordered magneto-ionic geometry during that event. 

Multi-wavelength follow-up and archival searches impose additional constraints. Optical observation reveals no convincing counterpart down to a deep limiting magnitude, disfavoring a luminous white-dwarf binary interpretation. Archival XMM--Newton observations from 2011 yielded no detectable X-ray counterpart, setting a 99\% upper limit on the 0.2--10~keV flux. Together with the reported Chandra detection of X-ray bursts and subsequent non-detection, this suggests that the source can enter a deep X-ray quiescent state and motivates rapid, triggered X-ray follow-up during future radio-bright windows.


\section{DISCUSSION}
DART J1832-0911 provides several important advances over previously reported LPTs. First, it is the strongest current case for a physical association between an LPT and a supernova remnant, thereby providing the clearest observational support so far that at least some isolated LPTs originate from young neutron stars. Second, the plausible SNR association supplies an external age constraint, allowing the ultra-long spin period to be discussed in an evolutionary context and making post-supernova braking scenarios, such as fallback-disk interaction, substantially more plausible. Third, the source exhibits distinctive polarization behaviour, including phase-locked circular polarization and a high linearly polarized dwarf pulse, which places new constraints on the radio-emission geometry and further distinguishes this source from LPTs associated with white-dwarf binary or accretion-powered systems. Taken together, these properties favor a young neutron star as one origin of at least a subset of long-period radio transients.

 The spatial association between this source and the SNR indicates that it is likely the stellar remains of a supernova explosion, specifically a neutron star rather than a white dwarf. Given the characteristic age of both DART J1832-0911 and SNR G22.7-0.2, it is reasonable to suggest that DART J1832-0911 might be a young neutron star. The emission properties further suggest a pulsar-like emitter of DART J1832-0911. The narrow, phase-locked CP indicates a stable geometric structure of the magnetic field of DART J1832-0911 even at small scales. This stability argues against models that rely on dynamic magnetic fields as the primary driver of radio emission, such as star flares \cite{benz2010physical} and dwarf binary systems \cite{marsh2016radio,de2025sporadic}.  The nearly 100\% linearly polarized burst signals support both theoretical radio emission mechanism and observation related to magnetars\cite{camilo2006transient,camilo2007polarized,kramer2007polarized,levin2012radio}, which is believed to be a class of young neutron stars \cite{kaspi2017magnetars}. Such short-lived pulse emitted during an emission-dull or -quiescent state is similar to the dwarf pulse observed during pulsars' nulling period\cite{chen2023strong,yan2024dwarf,srostlik2005core,kloumann2010long,esamdin2012psr}. These emissions are believed to arise from rain-like drops of particles formed via pair production near the neutron star surface\cite{chen2023strong,yan2024dwarf}. It is worth noting, however, that in the case of LPTs, such emissions may occur regularly during both emission-loud and -quiet phases, but remain undetected due to the sensitivity and temporal resolution limitations of DART.

A neutron star assumption of LPTs significantly challenges conventional evolution models of fast spinning neutron star scenarios\cite{rea2024long}, which typically have rotation periods of less than a few minutes, as illustrated in Fig.\ref{ppdot}. Theoretically,  neutron stars---formed as remnants of supernova explosions---are born with fast spin periods on the order of milliseconds if they inherit most or all of the angular momentum from their progenitors. Consequently, the emergence of LPTs suggests that these stars could form under two possible conditions: either with a significant loss of angular momentum from their progenitors such as neutrino emission \cite{epstein1978neutrino,camelio2016spin}, or through some rotational braking process during their spin evolution (see \cite{heger2005presupernova} for a review). 

Given that DART J1832$-$0911 is embedded in the SNR environment, rotational braking by a supernova fallback
disk \cite{menou2001stability,ronchi2022long,yang2024instability} provides the most natural explanation for its ultra-long spin period.
We simulate the reasonable period range based on a fallback disc model \cite{yang2024instability}, as shown in Fig.\ref{ppdot}. The results suggest that the observed LPTs could be well explained in the period between $10^{4}$ and $10^{5}$ years after the supernova explosion. Therefore, the association between DART J1832-0911 and SNR G22.7-0.2, with an age estimate ranging from $3\times10^{4}$ to $2\times10^{5}$ years, could further support the hypothesis that such LPTs may originate from interactions between young neutron stars and fallback discs.

~\\

\begin{acknowledgments}
This work is supported by the Strategic Priority Research Program of the Chinese Academy of Sciences. 
\end{acknowledgments}

\clearpage
\setcounter{figure}{0}
\setcounter{table}{0}
\setcounter{equation}{0}
\setcounter{section}{0}
\renewcommand*{\thefigure}{S\arabic{figure}}
\renewcommand*{\thetable}{S\arabic{table}}
\renewcommand*{\theequation}{S\arabic{equation}}
\renewcommand*{\thesection}{S\arabic{section}}
\newcommand{\titlefont}{\fontsize{14}{16}\selectfont\bfseries}
{\titlefont\centering Supplementary data for\\ A 44-minute periodic radio transient in a supernova remnant \par}

\section{DART Observation \& Data Reduction}
\label{dart_obs}
\hfill
\\
DART (previously known as DSRT, for a detailed description, see Yan et al. 2023 \cite{yan2023super}) is a synthetic aperture circular array telescope completed in 2023 for dynamic solar imaging. Located in Daocheng, Sichuan Province, China, DART is part of the Chinese Meridian Project’s key observation sites \cite{wang2024china}. The array consists of 313 parabolic antennas, each with a 6-meter aperture evenly distributed on a circle, forming a configuration with a diameter of 1 kilometer. We utilize DART to image the sky in the frequency band of 149-459~MHz. The maximum field of view (FOV) of DART at 150 MHz is approximately $20^{\circ}\times20^{\circ}$. For the high-frequency band, the angular resolution is approximately 100 arc-seconds. By fitting the point spread function (PSF) of a bright point source within a 3 dB angular width, DART achieves a positional precision of a few arcseconds for point sources. The telescope’s ability to produce continuous interferometric images with a temporal resolution as fast as 5 milliseconds makes it an exceptionally powerful tool for surveying dynamic phenomena across the Universe.

The DART backend simultaneously records interferometric visibilities at two distinct frequency channels, each with a maximum bandwidth of 4 MHz. The central frequencies of the two channels can be independently selected anywhere within the full receiving band of 150--450~MHz, with a maximum separation of 40~MHz between the two channels. All the frequencies mentioned refer to the mid-frequencies of two distinct 4 MHz-wide bands. The DART backend records visibilities at these two central frequencies without further channelization — each visibility set corresponds to one image at a single mid-frequency. Therefore, the frequency resolution of the visibilities is effectively the full 4 MHz bandwidth per channel. Consequently, multiple epochs at different central frequencies retrieve the frequency-dependent effect of dispersion. 

\begin{figure}
  \centering
  \includegraphics[width=0.95\linewidth]{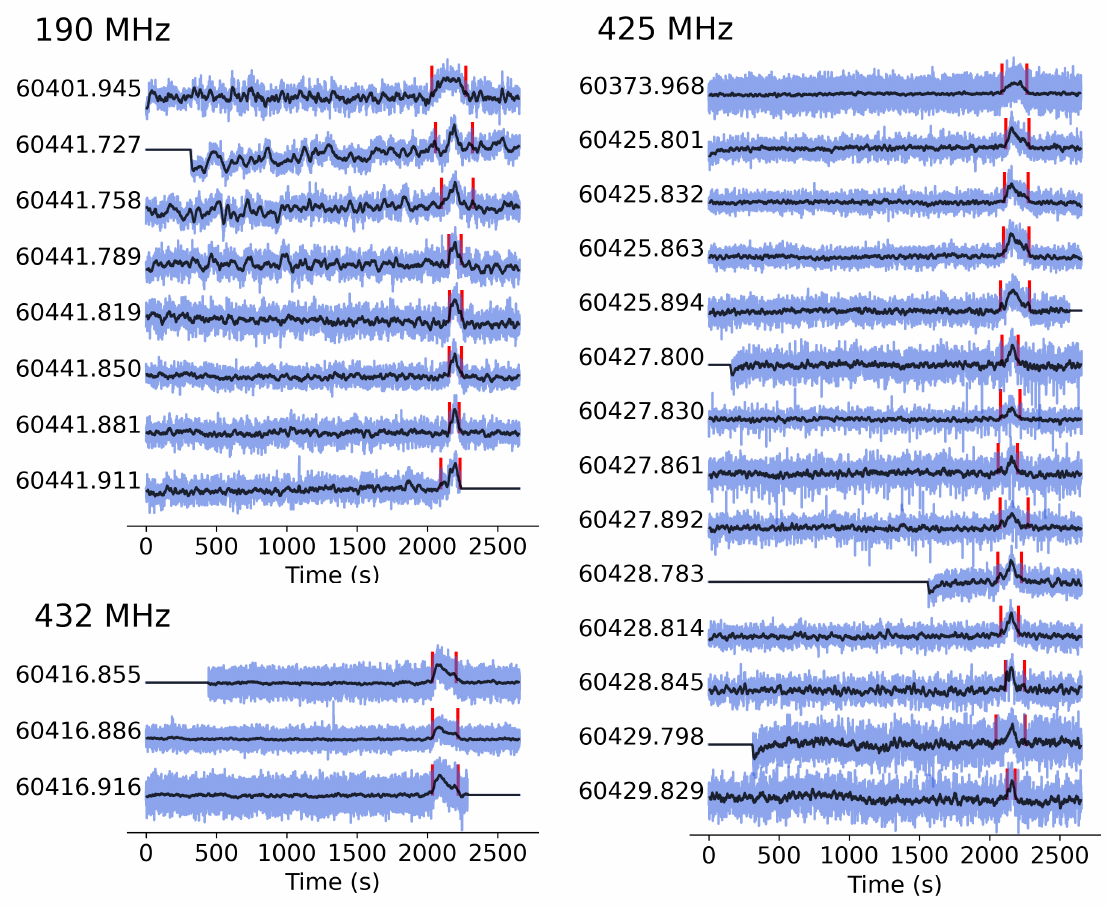}
  \caption{\textbf{Part of single pulse profile collected by DART.} We present a portion of the single pulse at 190~MHz, 425~MHz, and 432~MHz. The blue profiles are original samplings with different temporal resolutions, which are 100~ms (MJD~60373.968), 800~ms (MJD~60401.945), and 500~ms (others). The red profiles are smoothed series by running mean process.}
 \label{profile}
\end{figure}

For the majority of the data, we configure the integration time within a range from 100 to 500 milliseconds. For each observation, two center frequencies are selected from within the entire frequency band. Ultimately, a pulse series spanning from 190 MHz to 455 MHz is recorded. All these pulses are acquired with complete Stokes components.  Fig.\ref{profile} showcases a portion of the pulse profile components.

\begin{figure*}
  \centering
  \includegraphics[width=0.8\linewidth]{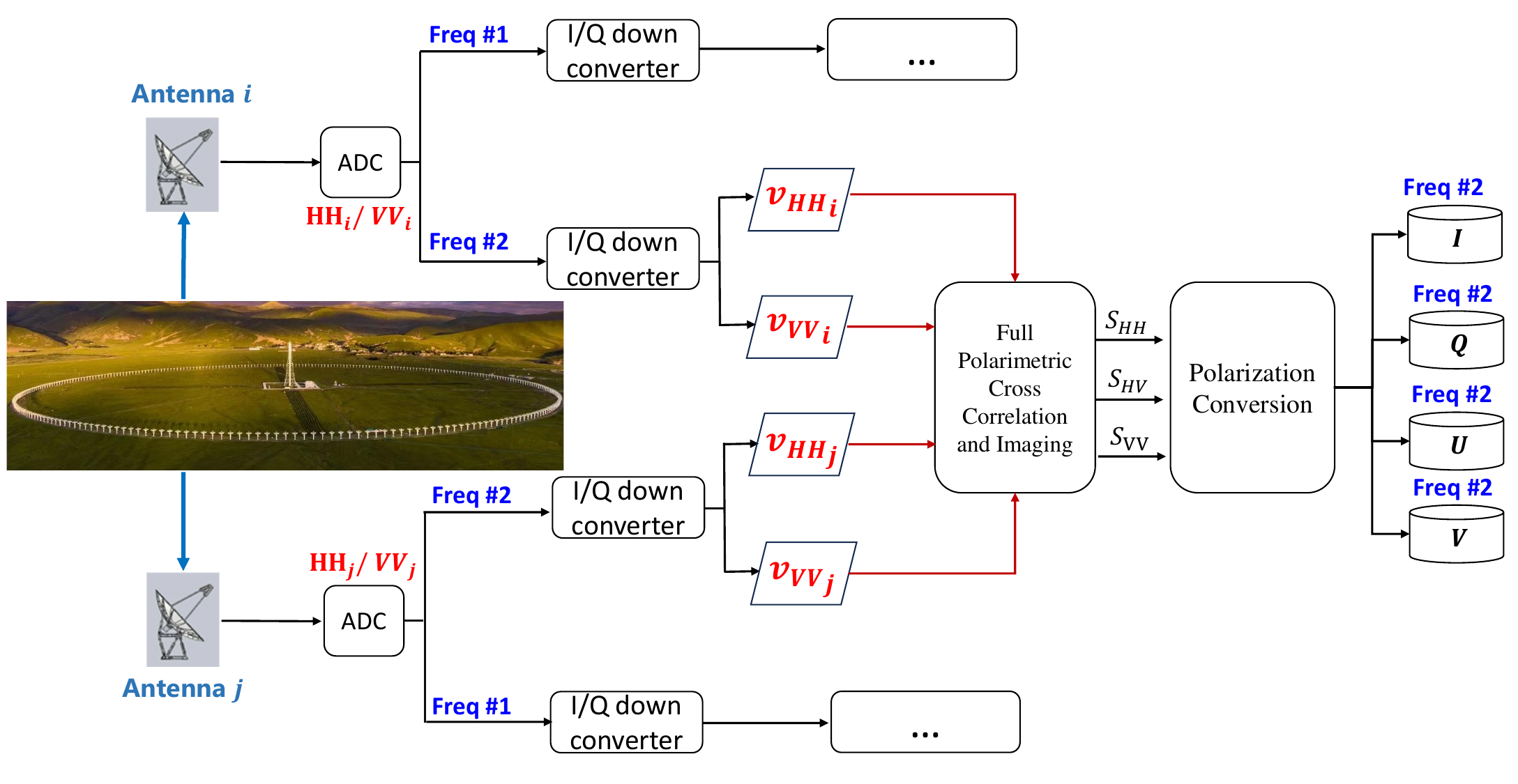}
  \caption{\textbf{Data processing pipeline of DART.} The raw data was recorded with full polarization at dual frequency. All samples of the continuous visibility function $v_{HH/VV}$ from 626 links (comprising 313 antennas with two orthogonal polarizations) will undergo independent cross-correlation and imaging processing at each of the two frequencies.}
 \label{pipeline}
\end{figure*}

In DART system, the polarimetric measurement accuracy of DART is achieved by amplitude and phase unbalance calibration with the external and internal calibration.  The calibration refers to the correction of instrumental gain and phase mismatches between all 626 channels.  Specifically, the measurement is composed by direct dual-pol observation(HH,VV) and full-polarimetric cross-correlation (HH,VV and HV). 
A post-data processing can finally convert the polarimetric measurement to the Stokes parameters as the form like
\begin{equation}
\begin{aligned}
I &= \frac{S_{HH}+S_{VV}}{2},\\
Q &= \frac{S_{HH}-S_{VV}}{2},\\
U &= Re(S_{HV}),\\
V &= -Im(S_{HV}).
\end{aligned}
\end{equation}
We illustrate the data processing chain in  Fig.\ref{pipeline}.
The polarimetric measurement accuracy only depends on the amplitude and phase unbalance of the DART array. Polarimetric calibration of DART is applied by the channel-based systematic calibration. With the external and internal systematic calibration, the amplitude and phase accuracy of 0.2~dB and $2.5^{\circ}$ between any pair of receivers is guaranteed. With the low residual calibration errors, the polarimetric measurement error of DART is lower than 5\% \cite{yan2023super}.

\section{FAST Observation and Flux Estimation}
\hfill
\\
We conducted a joint observation of DART and FAST at MJD 60559, with the integration time at 1 hour. The FAST frequency band and sampling time are 1-1.5~GHz and 196~us. 
The 19-beam receiver of FAST adopts orthogonal linear polarisation feeds, the diode noise signal will be injected as the polarization calibrated reference before the target signal recording\cite{li2018fast,jiang2019commissioning,jiang2020fundamental}. We conduct the polarimetric calibration by using the \rm{PSRCHIVE} software. 

We estimated the flux density of the pulse based on its SNR, using the radiometer equation:
\begin{equation}
    f_v=\frac{T_{\textrm{sys}}\cdot SNR}{G\sqrt{2B\Delta t}},
\end{equation}
where $T_{\textrm{sys}}\simeq24$~K and $G\simeq16$~K/Jy represent the  system temperature and gain of FAST \cite{jiang2020fundamental}, respectively. $B,t$ are the bandwidth and temporal resolution of the observation.
\section{Parkes UWL Observation}
\hfill
\\
We conducted a joint observation of DART and Parkes at MJD 60425.
We used Parkes to observe toward the initial position (RA = 18:32:45.0, Dec = -09:12:00.0) for 180 minutes starting from MJD 60425.74625, during which the transient is on the radio-active epoch. We used the standard Parkes backend, Medusa, which applies automatic gain correction for each sub-integration (in a timescale of 0.524288 s). We restored the original result using the scale and offset value recorded in PSRFITS. We extract Parkes UWL subband 0 to 2 (frequency 704 -- 1088 MHz) and averaged the entire 384 MHz. We used a time resolution of 0.524288 s, and applied Savitzky-Golay smooth with second-order polynomial fit and a window length of 51 time-samples. 

We confirmed that no pulses with an SNR greater than 3 were detected, whereas DART was able to detect them. We contend that the failure to detect these pulses is due to the automatic baseline correction employed by the backend of Parkes.

\section{GTC Observation}
\hfill
\\
We obtained the \textit{r}-band image of the DART source using the broad-band imaging mode of the OSIRIS$+$ on the Gran Telescopio CANARIAS (GTC) on 30 July 2024. 
The image was co-added using 24 single exposures of 150 s each to avoid sky saturation. 
A four-point dithering was used to avoid cosmic ray contamination. 
The total on-source time was 3600 s, and the observation duration lasted about 1.25 hr, including overheads. 
The observations were conducted under good weather conditions (seeing$\sim0.7^{\prime \prime}$). 
The data were first bias-subtracted, flat-fielded, cleaned of cosmic rays, and combined using the \textit{ccdproc} \cite{craig2017astropy} and \textit{DrizzlePac}\cite{hoffmann2021new} packages following standard techniques. 
The sky background of the entire image was then modeled and subtracted, primarily using the \textit{photutils}\cite{bradley2016photutils} package, following the instructions for modeling the complex 2D background in the Cross-Instrument section of the JWST Data Analysis Tool Notebooks\footnote{\url{https://spacetelescope.github.io/jdat_notebooks/notebooks/cross_instrument/background_estimation_imaging/Imaging_Sky_Background_Estimation.html}}.
The $3\sigma$ detection limit of the co-added image is 24.8 mag, while the $1\sigma$ upper limit of the background is 26.6 mag.

We searched for sources within a 5$^{\prime\prime}$ radius around the DART coordinates by performing aperture photometry on the co-added image using a 2$^{\prime\prime}$ radius.
The top panel in Fig~\ref{opt} displays the four sources closest to the DART coordinates, along with the reference star.
The light curves of these five sources are shown in the bottom, where $\Delta$mag represents the difference between the photometry of each individual exposure and its mean magnitude during the observations.
Details of the five targets are provided in Table~\ref{tab:nearbyS}.
\begin{table}[htb]
    \centering
    \caption{The separations between the nearby sources and the central coordinates of the DART location, as well as their \textit{r}-band magnitudes.}
    \medskip
    \begin{tabular}{lcc}
    \hline
    \hline
    \noalign{\smallskip}
    Source              & Separation & Magnitude \\
                           & (arcsec) & (mag) \\
    \noalign{\smallskip}
    \hline
    \noalign{\smallskip}
    s1              & 1.5 & 21.9 \\
    \noalign{\smallskip}
    s2               & 4.5 & 24.2 \\
    \noalign{\smallskip}
    s3               & 5.5 & 24.6 \\
    \noalign{\smallskip}
    s4               & 5.7 & 25.6 \\
    \noalign{\smallskip}
    Reference        & 8.1 & 20.1 \\
    \noalign{\smallskip}
    \hline
    \end{tabular}
    \label{tab:nearbyS}
\end{table}

Our analysis indicates that no optical counterparts are likely associated with DART J1832-0911. Source $s1$ shows no significant variability during the observation period. Cross-matching sources $s1$ through $s4$ with archival data reveals that only $s1$ is detected in Pan-STARRS DR2 \cite{magnier2020pan}, exhibiting a comparable $r$-band magnitude. This suggests that $s1$ is a typical main-sequence star.  Sources $s2$ to $s4$ exhibit variability in their light curves. However, $s4$ is too faint to be reliably detected in individual exposures, and its apparent fluctuations are likely due to detection uncertainties. Moreover, $s3,s4$ lies outside the plausible location range of the radio transient, further weakening its candidacy. The variable light curve of $s2$ is characteristic of pulsating variable stars, which are known to produce thermal rather than non-thermal radio emissions \cite{christy1966pulsation}. As a result, none of the optical sources examined appear to be viable counterparts to DART J1832-0911.

The optical non-detection at the DART coordinates sets an upper limit on its optical counterparts. 
The radio emissions might originate from a faint dwarf star below our detection threshold, or from other objects not expected to be visible in optical observations, such as a neutron star.
In the case of a faint white dwarf, assuming an upper limit of 26.6 mag and a white dwarf radius of $r=0.2 \rm R_{\odot}$, the corresponding effective temperature would be $\rm T_{eff} \sim 17000 K$ or lower. 

The GTC observations place strong constraints on the presence of a luminous dwarf companion of DART J1832-0911 at the estimated distance of ~4.5 kpc (with uncertainty of 1.5 kpc). However, we cannot completely rule out the existence of a very faint or cool dwarf below the detection threshold.

\section{Timing Analysis}
\hfill
\\
We collected more than 200 single pulses in the 150–450 MHz range, among which 28 pulses at 425 MHz are with high signal-to-noise ratios. We derive the period through standard pulsar timing methods. Before the timing procedure, we had transferred the local time (in MJD) to the barycenter frame of the solar system. We first use a prior period of 2656.2~s to adopt the single pulses from a continuous time series. Then, we select several pulses with high SNR and take their mean profile as the standard template, to cross-correlate any other pulses to determine the TOA for each pulse. Specifically, we calculated a series of correlations between any single pulse and the standard profile by shifting the bins (i.e. the TOAs) of single pulses. Then we fitted the relation of correlations versus TOAs by a Gaussian function and determined the proper TOA based on fitting result. The TOA errors are derived at a $3~\sigma$ level of Gaussian fitting rather than the conventional 
$1\sigma$, reflecting the fundamental limitations of our timing approach: the time series are extracted from dirty image cubes with a 4~MHz bandwidth and no intra-channel dedispersion, and the finest time resolution is 100~ms against pulse widths of order 100~s. Under these conditions, the $3\sigma$ threshold provides a more conservative and robust estimate of the TOA uncertainty than the standard $1\sigma$ 
convention used in high-precision pulsar timing. Given the large TOA uncertainties inherent to the imaging-based time series and the absence of intra-channel dedispersion, a reliable DM fit from multi-frequency TOAs is not feasible with the DART data alone. We therefore rely on the FAST observation for a precise DM determination, as described in 
Section~S7. 

\begin{figure}
  \centering
  \includegraphics[width=1\linewidth]{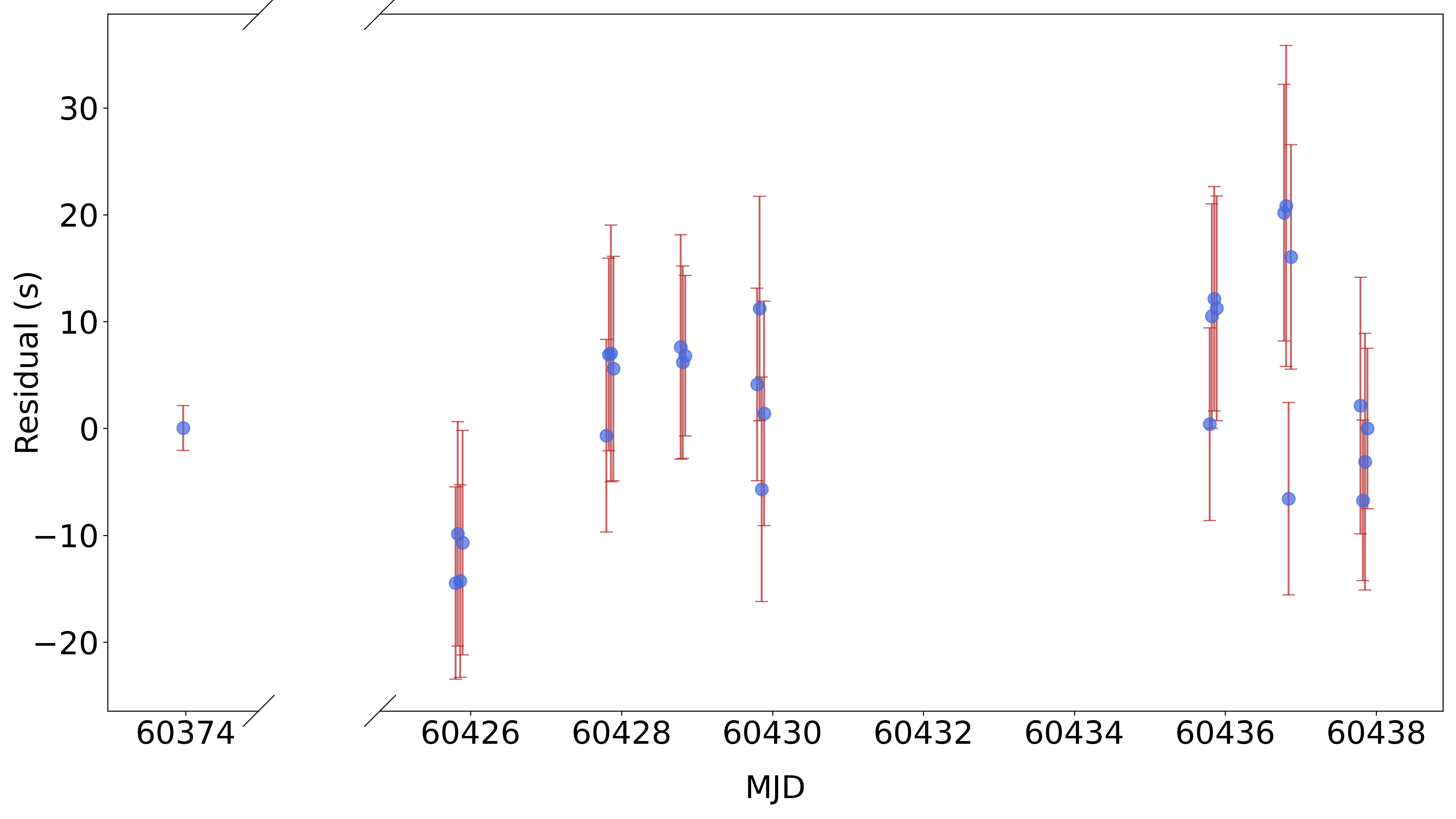}
 \caption{\textbf{Timing analysis of DART J1832-0911}. The pulse sequences are recorded at 425 MHz and over a span of approximately 64 days. }
 \label{timing}
\end{figure}

We collect a group of TOAs covering more than two months, a long nulling state prevents us from recording more data. We use the least-square timing method to fit TOAs and derive a period $P$ at $2656.23\pm0.15$~s. The TOA fitting result is shown in Fig.\ref{timing}. The period derivative we estimate is $\dot{P} < 5\times 10^{-9}$~ s/s. It is important to mention that the fitted TOAs belong to pulses at a central frequency of 425~MHz,  because we collected the largest number of pulses at this frequency. Fig.\ref{ppdot} shows the P-Pdot diagram of DART J1832-0911. We also present another three reported LPTs in the figure. 

\section{Location Constrains and Surrounding Compact Objects}
\label{subsec:ra_dec}
\hfill

\begin{figure*}
  \centering
  \includegraphics[width=0.8\linewidth]{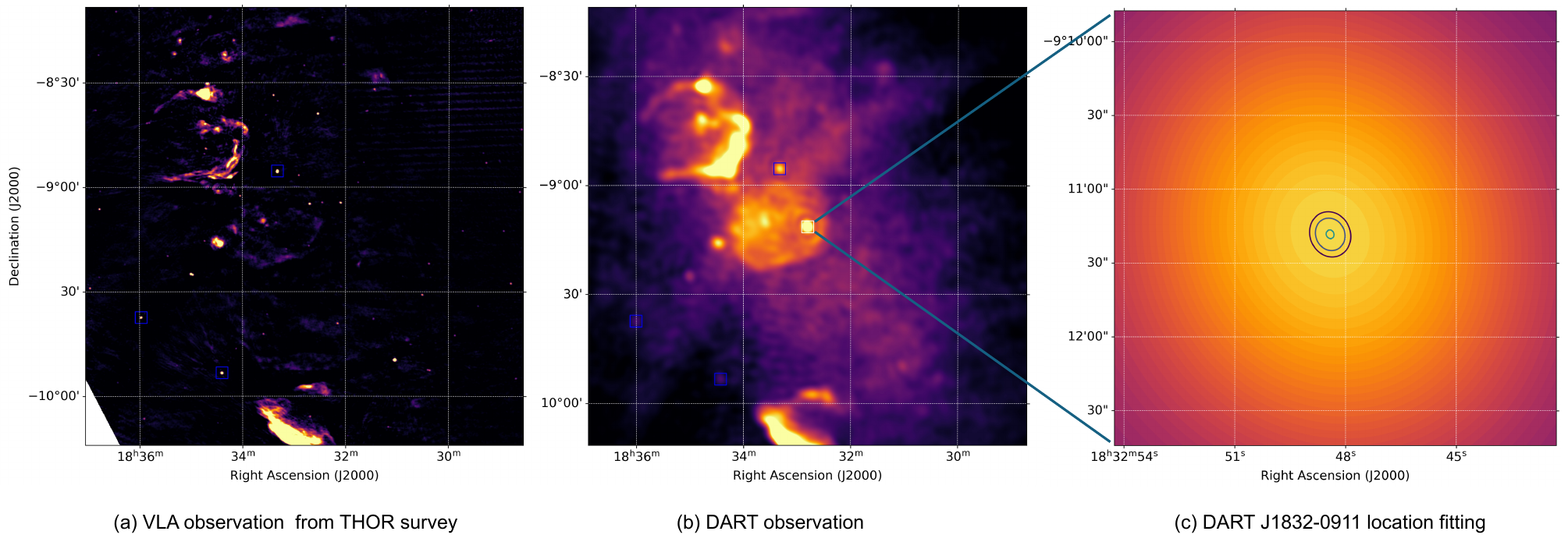}
 \caption{\textbf{The location projection of the DART observation modulated by the VLA observation.} The blue squares mark the reference sources common to both observations. (a) VLA observations; (b) DART observations; (c) The localization of DART J1832-0911, where the contour lines represent confidence levels of 68.3\%, 95.5\%, and 99.7\% from inner to outer regions, respectively. We adopt the center of the 99.7\% confidence level as the final source position: RA = 18h32m48.5s and DEC = $-$09d11m17s, with an uncertainty at $\pm5^{\prime\prime}$.}
 \label{loc}
\end{figure*}

We employ a super-resolution technique to achieve a beam size of 0.5 arcsecond, applying zero-padding interpolation to the sparsely sampled $(u, v)$-plane of DART observations. 
To mitigate systematic errors and phase biases induced by atmospheric turbulence, we align the DART image coordinates with the VLA reference frame using observation from the VLA THOR survey at 1.06 GHz \citep{bihr2016continuum}.
We select three point sources that can be distinguished from the Galactic radio background in both DART and VLA observations, as marked by blue squares in Fig. \ref{loc}. The DART coordinate frame is calibrated by fitting these reference points to the VLA frame 
using the ``\texttt{fit\_wcs\_from\_points}" function from the  \texttt{astropy} package. The resulting super-resolution DART images, presented in panel (b) of Fig. \ref{loc}, were obtained at 414 MHz with an integration time of 1,000~s, capturing the LPT emission shown in panel (c).

For precise localization of the target source, we perform a two-dimensional Gaussian fit to the emission region, defined by a 10\% (or -0.1~dB) decrease from the peak intensity of the main beam. The fitting results, illustrated in panel (c) of Fig. \ref{loc}, include contours representing the 68.3\%, 95.5\%, and 99.7\% confidence levels, 
progressing from innermost to outermost regions. The final estimated position of the target source is  RA = 18h32m48.5s, Dec = $-09^\circ11'17''$, with an uncertainty of $\pm 5^{\prime\prime}$ at the 99.7\% confidence level.

Taking the age of SNR G22.7-0.2, we can estimate the escaping velocity of DART J1832-0911 if it was born in the center of this SNR. The 2D transfer speed of a pulsar can be estimated by $V_T = 4.74\mu_{total}D_{dis}$\cite{lyne1994high}, where $\mu_{total} = \sqrt{(\Delta \alpha \textrm{cos}\delta)^{2} + (\Delta\delta)^{2}}$, $\alpha$ and $\delta$ are the Right Ascension and Declination in degrees. According to the age limits of the SNR, the 2D transfer velocity of DART J1832-0911 is 176 km/s for 30,000 years and 26 km/s for $2\times10^{5}$ years. This proper motion speed is quite acceptable in the range of the pulsar proper motion statistics \cite{lyne1994high,hobbs2005statistical}

DART J1832-0911 lies in the Galactic plane, a region rich with supernova remnants (SNRs) and compact objects. 
Within our $4^\circ \times 4^\circ$ field of view at 425 MHz, there are 7 identified SNRs \footnote{\url{http://snrcat.physics.umanitoba.ca}} and 39 known compact objects \footnote{\url{https://www.atnf.csiro.au/research/pulsar/psrcat/}}, including pulsars and high-energy sources. We investigate the character age and DM distance of these compact objects, as indicated in the  Fig.\ref{snr_psr}. Objects with DM distances below 2 kpc or above 8 kpc, or characteristic ages exceeding $10^6$ years, are classified as association-unlikely candidates. These thresholds account for a 100\% uncertainty in DM distance and a 1000\% difference between the SNR's evolutionary age and the pulsar's characteristic age, following previous studies \cite{gaensler2000large}. 

Spatial distribution analysis reveals that, within a $4^\circ \times 4^\circ$ region centered on DART J1832-0911, the number density of compact objects and SNR is 2.5 deg$^{-2}$ and 0.4 deg$^{-2}$, respectively. After excluding unlikely candidates and confirmed associations, the density of potentially associated compact objects drops to 1 deg$^{-2}$. These densities are significantly lower than in regions with possible coincidental SNR-pulsar overlaps, where compact objects are often suspected to be wandering stars from nearby SNRs. For instance, the PSR J1844-0346 region hosts 10 identified compact objects and at least 13 SNR candidates within approximately 1 deg$^{-2}$ \cite{amenomori2022measurement,kato2024result}. Considering the density of the potential associations and the separation distance of the DART source to the SNR center, the probability of an unrelated pulsar coincidentally wandering into the LPT region can be approximated by the fractional area $\pi r_{SNR-LPT}^2/\mathrm{deg} ^{2}$, where $r_{SNR-LPT}^2$ is the projected localization radius of the transient (approximately 3 arcminutes). Based on this approximation, the chance of a random pulsar coinciding with the SNR location of DART J1832-0911 is 0.7\%. 

To investigate any possible associated compact object with SNR G22.7-0.2, we define a distribution region centered on the SNR’s coordinates, assuming that the missing neutron star escapes with a typical proper motion velocity of 200 km s$^{-1}$ \cite{hobbs2005statistical}, and a time scale of the limited age of G22.7-0.2. This region, spanning a projected radius of 40~pc, is depicted as a white dotted circle in Fig.\ref{snr_psr}. 
There are 1 radio pulsar J1834-0915 \cite{ng2015high} and 1 $\gamma$-ray object HESS J1832-093\cite{hess2015discovery}. No specific study has been found about the radio pulsar, and research about the HESS J1832-093 strongly against its association with any SNRs \cite{hess2015discovery} due to its the spatial separation (a lower limited distance at 5~kpc and no alignment with the SNR shell) and a high-mass binary counterpart\cite{hess2015discovery,marti2020x}. 

Furthermore, no reported cases exist of pulsars initially discovered near supernova remnants (SNRs) being later confirmed as chance coincidences, according to prior PSR-SNR association studies. Although proper motion measurements of some pulsars, e.g. PSR B1757-24 \cite{hales2009proper,gaensler2000large},PSR J1709-4429 \cite{romani2005complex,de2021psr}, reveal a misalignment between the pulsar’s trajectory and the SNR’s center, an association remains plausible when accounting for unique explosion conditions \cite{de2021psr}, including the progenitor star’s wind, a significant birth kick, and interactions with the pulsar wind nebula (PWN) outflow.

Conclusively, we favor the DART J1832-0911 as the potential stellar remains of SNR G22.7-0.2 than any other compact objects. 

\section{Dispersion Measure (DM) and Rotation measure (RM)}
\label{subsec:dmrm}
\hfill
\\
We measured the DM from DART observation by calculating the time delay $\delta t$ between time series at two frequencies, 190~MHz and 205~MHz. The time series spans nearly 5 hours and includes 7 pulses, allowing us to establish a strong correlation between the two series. Nevertheless, the coarse time resolution (500~ms) and large pulse width prevent us from accurately measuring the delay. 
$\Delta t$ is determined by fitting the cross-correlation relations between pulse series at two frequencies. For each correlation coefficient, it was derived by shifting the low-frequency series forward step by step.  

\begin{figure}
  \centering
  \includegraphics[width=0.45\linewidth]{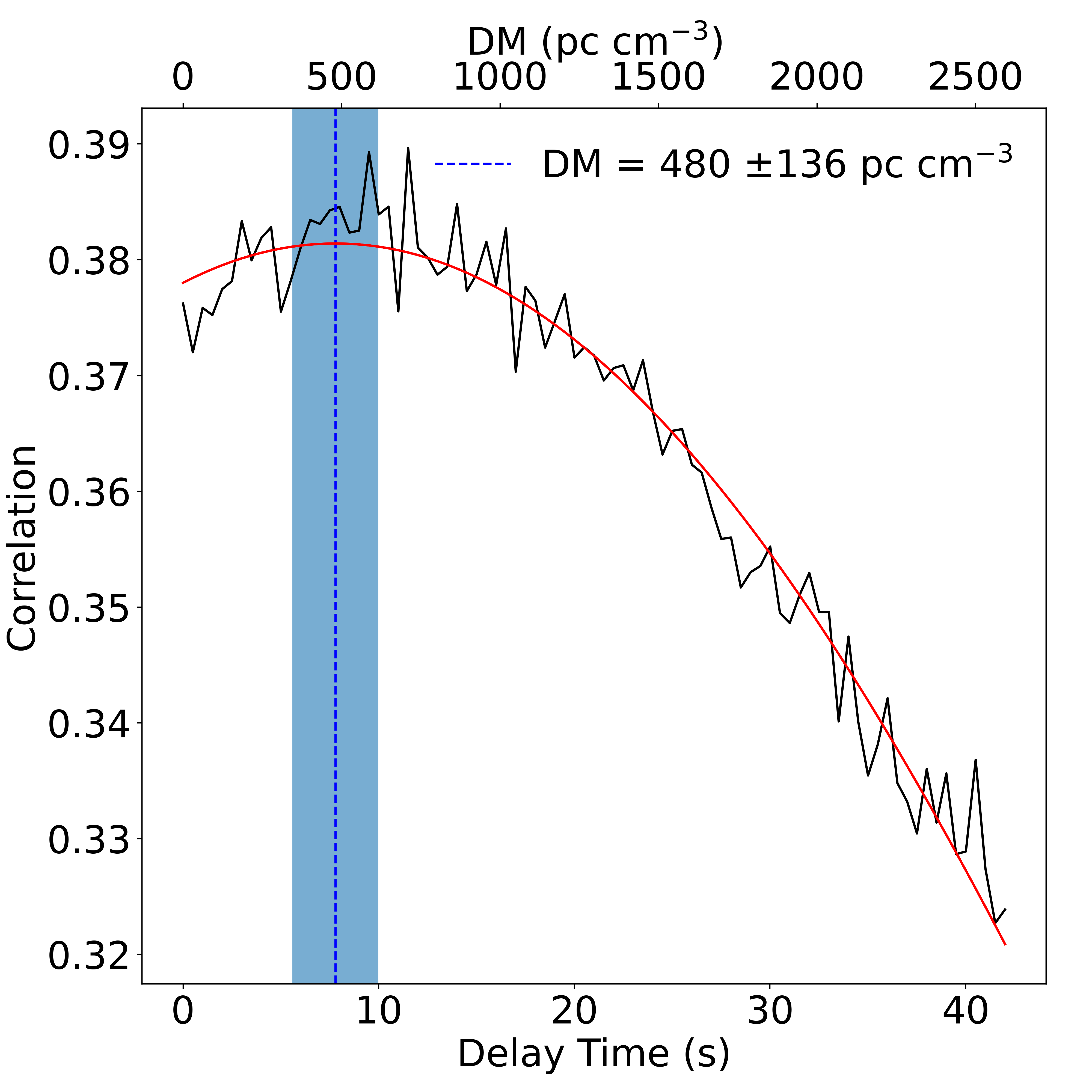}
   \includegraphics[width=0.45\linewidth]{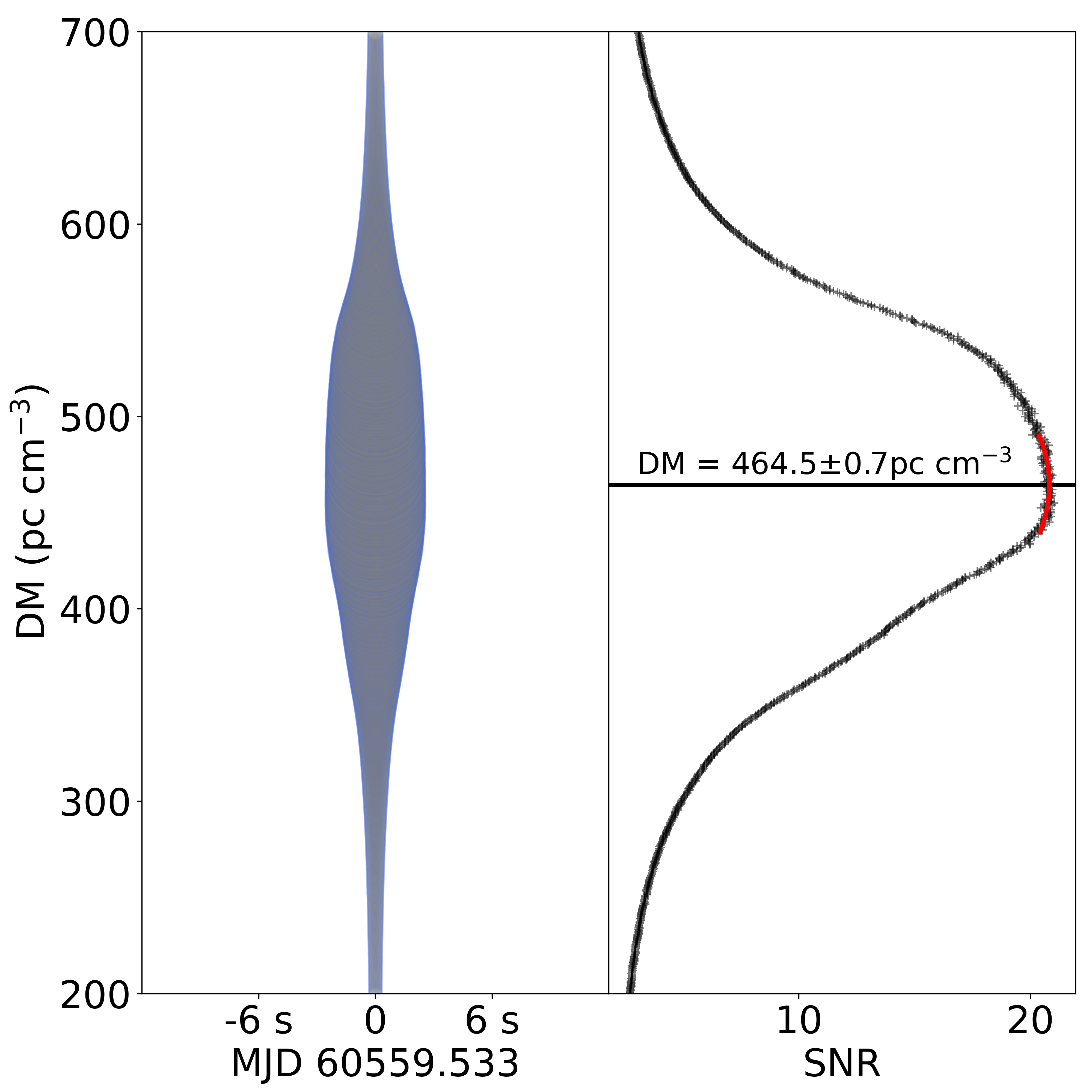}
\caption{\textbf{DM estimation based on data from DART (left) and FAST (right).} DART: The function of correlation coefficient (black line) against delay time (i.e. DM) between time series at two frequencies. The red line depicts the fitting relation, and the blue range covers a confidence interval within 3~$\sigma$. FAST: The single pulse search results.}
 \label{dm_fit}
\end{figure}

Using the dispersion relation $\Delta t=4.148808\times 10^{3}\cdot(1/f_{190 ~\textrm{MHz}}^{2}-1/f_{205~\textrm{MHz}}^{2})~\times$~DM, one can derive the DM value through the $\Delta t$, at which the two series are most correlated.   
Our estimation shows that the time delay between 205~MHz and 190~MHz could be as long as ten seconds, and the corresponding DM is 480$\pm$136~pc cm$^{-3}$. The left panel of  Fig.\ref{dm_fit} shows the correlation coefficients versus the delay time (or DM) and our fitting results.

We measured the DM from FAST spectral by fitting the SNR of de-dispersed pulse at different DMs. The right panel of  Fig.\ref{dm_fit} shows the SNR of pulse de-dispersed under different DMs.

We measure the RM based on the spectrum observed by FAST. By using the \rm{rmfit} tool of \rm{PSRCHIVE} \cite{hotan2004psrchive,van2012pulsar}, we got the RM estimation of $+105.6\pm 0.1 ~\textrm{rad}$~$\textrm{m}^{-2}$.

The DART data do not allow a robust, independent determination of the RM, because in the DART imaging mode each polarization measurement is integrated over a single 4~MHz-wide band without further intra-channel frequency resolution (Section~S1). Therefore, the frequency sampling is insufficient for a reliable fit of polarization position angle as a function of $\lambda^2$. Moreover, once Stokes $Q$ and $U$ have been averaged over the full 4~MHz band, the information required for channelized derotation is no longer available, so the effect of in-band Faraday rotation cannot be corrected in post-processing within the DART data themselves.

Nevertheless, the RM measured from the FAST burst provides a useful consistency check. 
Following the standard complex-polarization formalism for Faraday rotation \cite{burn1966depolarization,sokoloff1998depolarization,brentjens2005faraday}, we define a convenient band-averaged depolarization factor,
\[
\eta_{\rm RM} = \frac{L_{\rm obs}}{L_{\rm int}}
= \left|\frac{1}{\Delta \nu}\int_{\nu_0-\Delta\nu/2}^{\nu_0+\Delta\nu/2}
e^{2i\,{\rm RM}\,\lambda(\nu)^2}\,d\nu\right|,
\]
where $\eta_{\rm RM}$ is the fraction of intrinsic linear polarization that survives after averaging over the full channel bandwidth. Here the RM value is ${\rm RM}=+105.6~{\rm rad~m^{-2}}$ and the expected bandwidth is $\Delta\nu=4$~MHz.  For representative DART frequencies, we obtain $\eta_{\rm RM}\approx 0.023$ at 150~MHz, $0.090$ at 190~MHz, $0.106$ at 220~MHz, and $0.844$ at 425~MHz. These values correspond to very strong in-band depolarization at the low-frequency end of the DART band and still non-negligible depolarization near 425~MHz. Thus, while the DART data cannot yield a reliable RM, the observed scarcity of linear polarization is compatible with the substantial bandwidth depolarization expected from the RM measured by FAST.

\section{Emission Characters}
\hfill
\\
DART J1832-0911 displays multiple emission characteristics, which we identify as three distinct feature modes: wide-pulse, narrow-pulse, and nulling (or weak emission below DART's detection threshold) or dwarf-pulse modes. Our observations show no evidence of regular evolution between these features.

\begin{figure}
  \centering
  \includegraphics[width=1\linewidth]{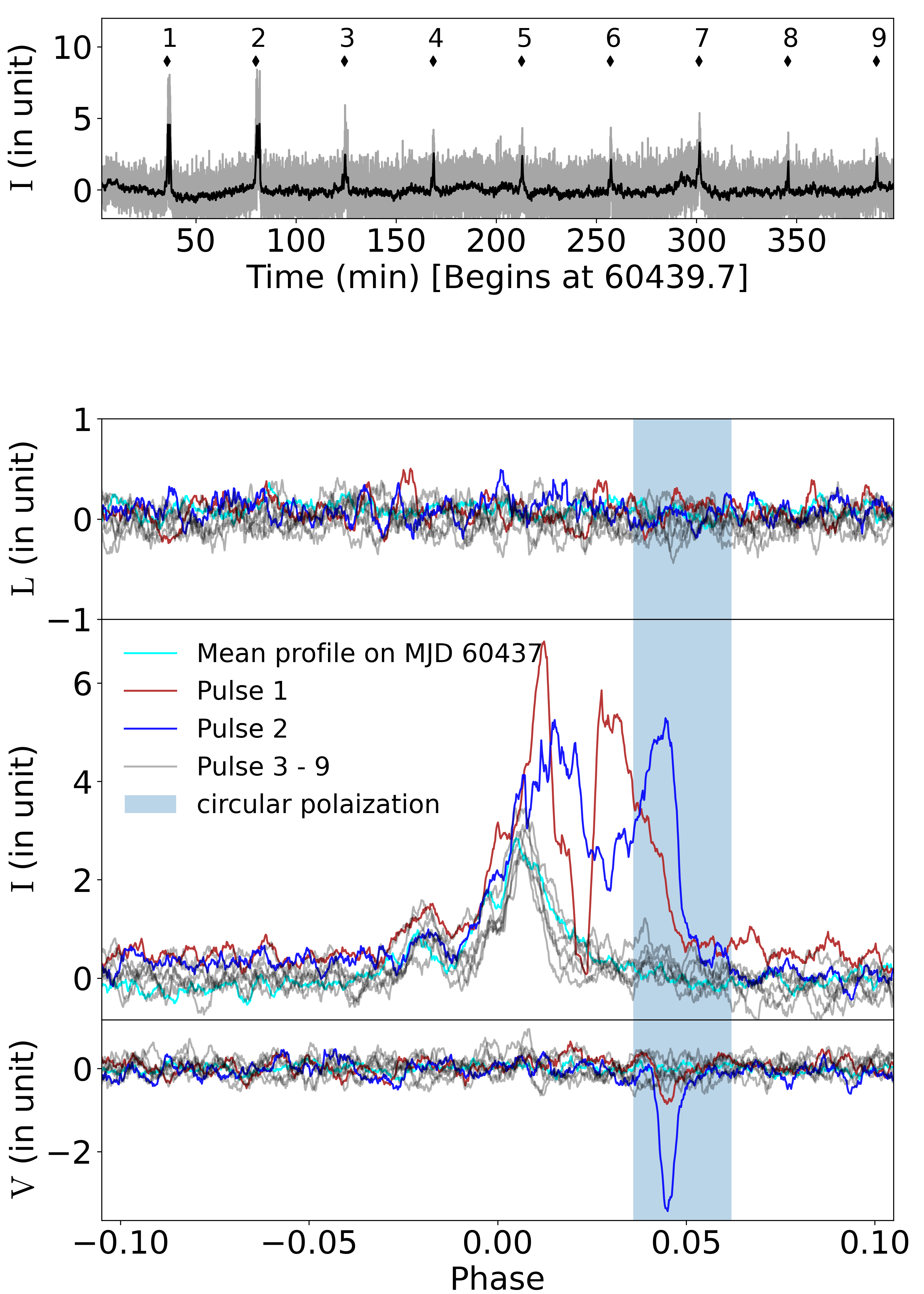}
 \caption{\textbf{A mode switching of DART J1832-0911 observed on 60439.7 (pulse 1-9).} The top panel is the time series of 9 continuous pulses. The three bottom panels are single pulse profiles of linear polarization (L), total intensity (I) and circular polarization (V).}
 \label{mode_change}
\end{figure}

In the wide-pulse mode, pulses are typically strong, with widths around 200--250~s, whereas in the narrow-pulse mode, the pulses are much weaker, with widths of roughly 40--100~s, as shown in the pulse stack figures Fig.\ref{supernova} and Fig.\ref{profile}.  
For most cases, the pulses remain consistent in both profile and strength, as shown in Fig.\ref{profile}.
However, we observed a significant disruption of this stability on MJD 60439, where flux and profiles dramatically changed, as shown in Fig.\ref{mode_change}. This sequence captures a transition from narrow-pulse to wide-pulse mode and back, with two strong wide pulses occurring within the narrow-pulse duration. Predictably, clear circular polarization components appeared at a fixed phase of the corresponding wide-mode pulse.
The temporal evolution of the emission---from strong, wide pulses to weak, narrow ones, and eventually to quiescence---suggests a dynamic process of particle accumulation and outward flow within the magnetosphere. 

The dwarf pulse of DART J1832-0911 (illustrated in Fig.\ref{fast}) appears to be similar to those observed in some pulsars during their nulling states \cite{chen2023strong,yan2024dwarf}. These sporadic, brief bursts 
are attributed to the emission from one or a few ``raindrops'' of particles generated through pair production in a fragile spark gap within the pulsar's magnetosphere, close to the neutron star's surface \cite{chen2023strong,yan2024dwarf}.
That is, a gap model is the underlying mechanism responsible for the radio emission characteristics of DART J1832-0911. Given the extended duration of the pulse components in the LPT, it is possible that such dwarf pulses may be prevalent throughout the entire radiation phase. 

\section{Polarization Characteristics}
\begin{figure}
  \centering
  \includegraphics[width=0.45\linewidth]{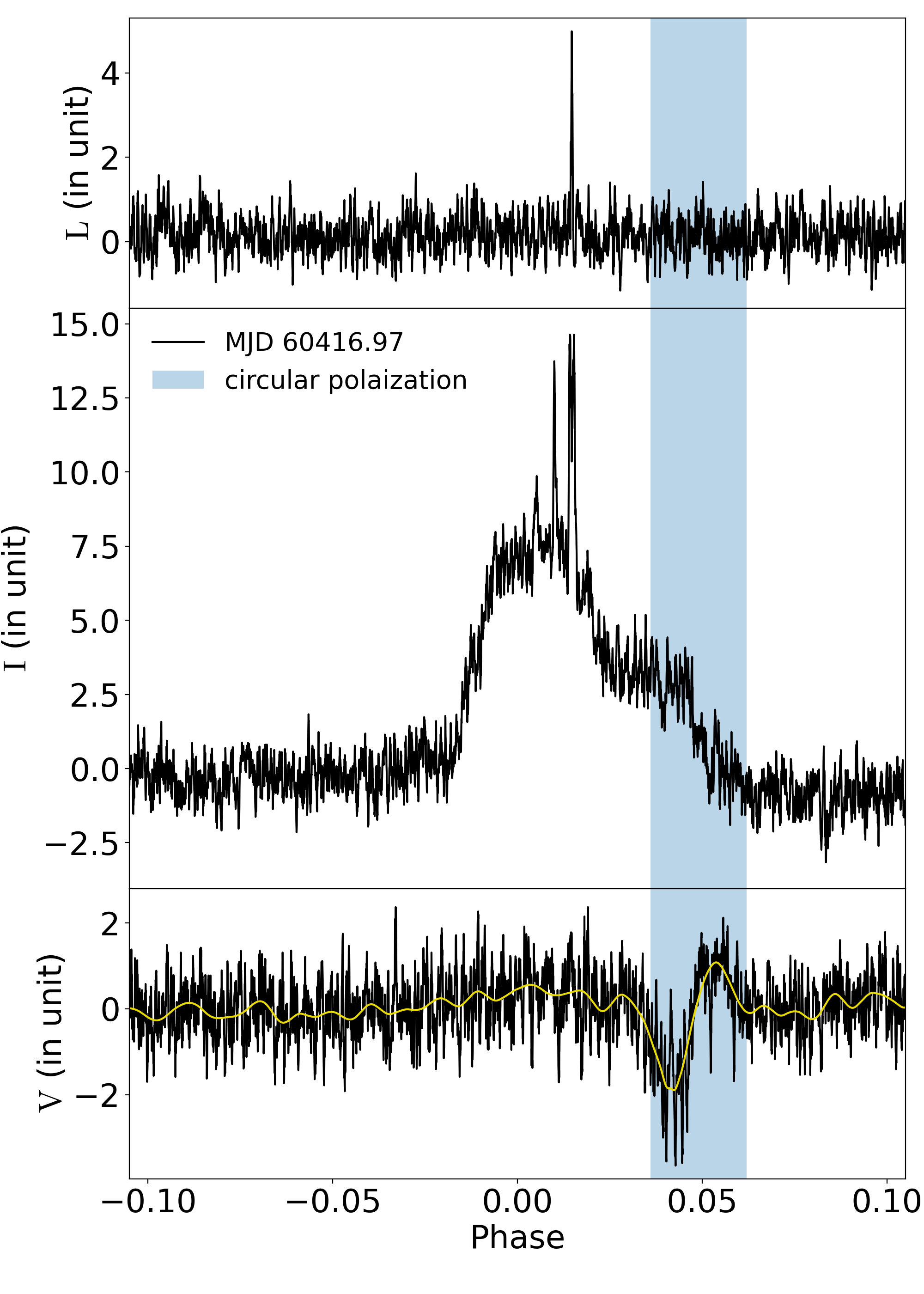}
  \includegraphics[width=0.45\linewidth]{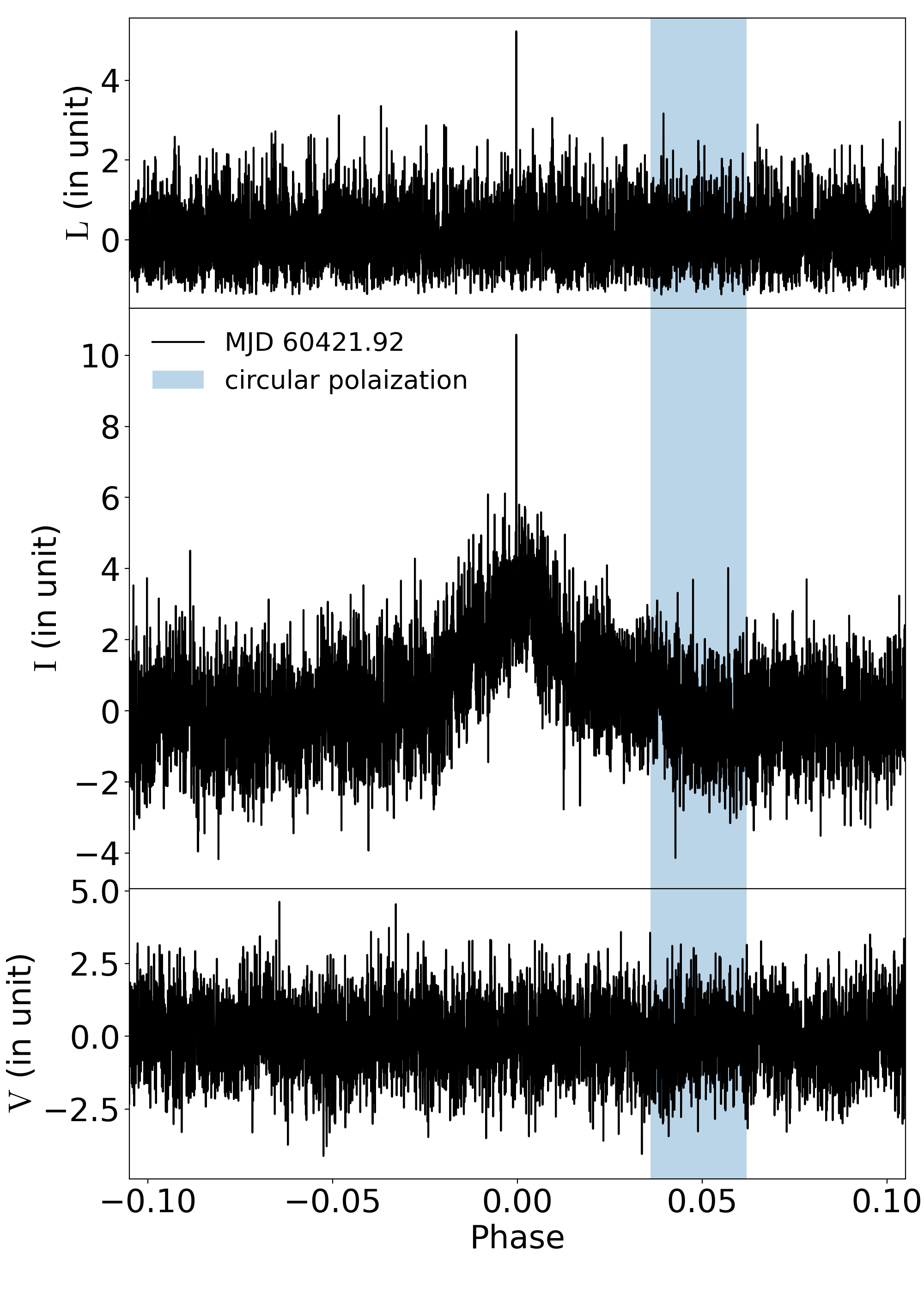}
 \caption{\textbf{The impulsive linear polarized burst (top panel) in two single pulses.} The temporal duration per bin for two pulses are 0.5~s for the left and 0.1~s for the right. Time series from the top to the bottom of each pulse are profile of LP,total intensity and CP.}
 \label{lp-burst}
\end{figure}

\hfill
\\
During the emission-active epoch observed from DART, we found poor polarization in most pulses. 
Our recordings show highly variable polarization, which includes: 
1, Absence of Polarization: Both linear and circular polarization were largely undetectable throughout most of the emission cycle. 2, Transient Linear Polarization: A linearly polarized signal occasionally appeared as a short burst.  Fig.\ref{lp-burst} shows two instances of these bursts, where linear polarization coincided with an intensity spike in the total signal. The majority of pulses, 
however, show no detectable linear polarization. 3, Phase-Locked Circular Polarization (CP): A stable, circularly polarized component appeared exclusively in the wide-pulse emission mode. Bottom panel of Fig.\ref{supernova} displays the phase-locked CP components recorded. Shaded areas in Fig.\ref{lp-burst} and  Fig.\ref{mode_change} highlight the CP profile ranges. These pulses include cases with reversed CP signs, cases with no sign change, and cases with no CP emission. These 

The scarcity of detectable linear polarization in the DART data is more likely attributable to observational limitations and propagation effects than to intrinsically unpolarized emission. At the low-frequency end of the DART band, the expected RM smearing within a single 4~MHz channel is severe enough to suppress most of the intrinsic linear polarization after frequency averaging. Even near 425~MHz, where the surviving linear-polarization fraction is higher ($\eta_{\rm RM}\approx 0.84$ for the FAST RM), the remaining signal can still fall below the detection threshold once limited sensitivity is taken into account. The typical noise level of the DART polarization time series is $\sim 100$~mJy; therefore, for narrow-mode pulses with peak flux densities of $\sim 200$--$400$~mJy, an intrinsic linear-polarization fraction of $\sim 10\%$ would correspond to only $\sim 20$--$40$~mJy in polarized flux density, well below the noise floor. In addition, the lack of intra-channel dedispersion broadens intrinsically short polarized components and further reduces their apparent peak signal-to-noise ratio. We therefore conclude that the rarity of linear-polarization detections in DART can be naturally explained by the combined effects of bandwidth depolarization, limited sensitivity, and temporal smearing. RM smearing is therefore likely the dominant limitation at the low-frequency end of the DART band, whereas at 425~MHz it remains an important, but not exclusive, contributor. 

During the emission-quiescent epoch observed with DART, we detected a bright burst due to the high sensitivity of FAST, with an estimated strength of 94 mJy , which is below DART's detection threshold. As shown in Fig.\ref{fast},  this pulse was highly linearly polarized, with linear and circular polarization degrees of 94\% and 3\%, respectively, and a flat position angle (PA) across the burst duration.

Our analysis supports an intrinsic properties of the phase-locked CP, suggesting that an intrinsic spin-down period rather than any extrinsic modulation consequence, e.g. an eclipse influence in a binary system and polarization conversion caused by SNR. During our observation period, both CP-active and CP-quiescent states persisted for several days, as did the wide- and narrow-pulse modes. Known binary systems are unlikely to remain in eclipse for multiple days. Furthermore, as shown in  Fig.\ref{mode_change}, CP appeared with burst signals even during the prolonged polarization-quiescent period, implying no strict periodicity in the CP-active phases.
The phase-locked CP may come from an open-field line region within magnetosphere of a pulsar.
The sign change of CP is attributed to the magnetosphere emissions since the line of sight sweeps across the minimum angle between the line of sight and the magnetic axis.

The high LP burst emitted during the DART's non-detection period demonstrates a strong magnetic field of DART J1832-0911, such near 100\% LP emission usually happens in some extreme emission phenomenons, like fast radio bursts (FRBs) ( e.g. \cite{michilli2018extreme,andersen2019chime,hilmarsson2021polarization}), and some of the giant pulse from radio pulsars \cite{ jessner2010giant,hankins2016crab}. Extremely high linear polarization places severe constraints on the radiation mechanism, only strongly magnetized sources can support relativistic particle bunches that emit highly linearly polarized signals. In addition, a flat PA, similar to some FRBs\cite{michilli2018extreme}, indicates an intrinsic limited duration of the emitting process, rather than being influenced by such geometric effects, such as the alignment between the rotation axis and the magnetic axis. One possible emission scenario is that plasma with insufficient density produces nearly 100\% linearly polarized emissions with a frozen PA, propagating from the strongly magnetized stellar surface to the exit point, where the wave undergoes minimal rotation. This process occurs on a short-lived timescale \cite{lu2019fast}.

\section{Model Constraints}
\begin{figure}
  \centering
  \includegraphics[width=0.95\linewidth]{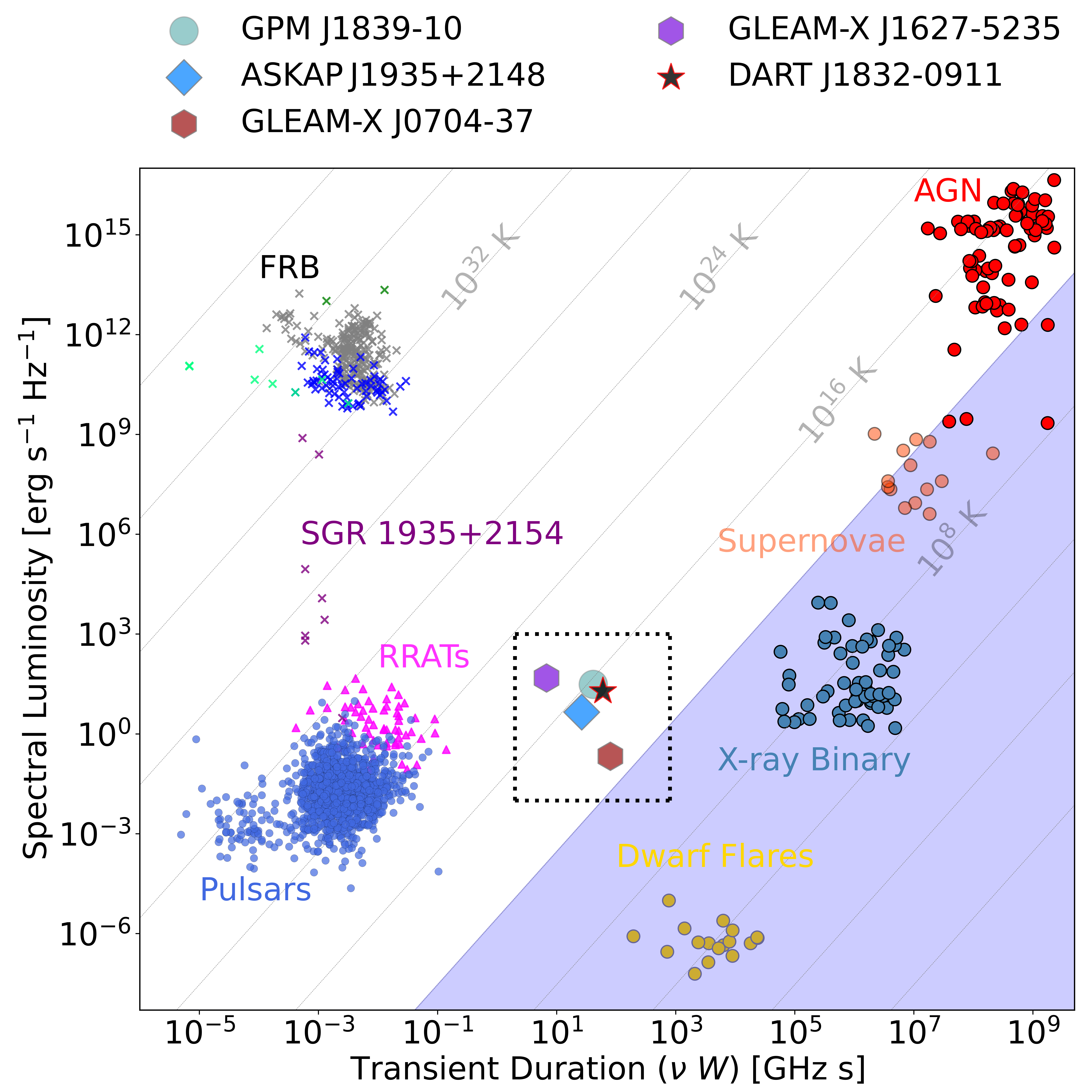}
  \caption{\textbf{Transient phase space diagram as a function of spectral luminosity against emission duration and frequency.} The LPTs cluster is marked in the black dashed box, of which the flux density and pulse width for the scatters of each LPTs are from the mean values provided from \cite{hurley2022radio,hurley2023long,caleb2024emission,dong2024discovery}. Grey diagonal lines indicate a limited brightness temperatures, which use $cw$ to represent the region size in Equation.\ref{eq2}. }
 \label{phase_dig}
\end{figure}

\hfill
\\
In addition to the directly observed proofs mentioned above, we discuss derived constraints about the nature of DART J1832-0911 in this section. According to the radio luminosity and emission duration, all reported LPTs are driven by coherent emission, same as transients like radio pulsars and FRBs, as shown in  Fig.\ref{phase_dig}. However, emission from WD binary system ILT J1101+5521 is predicted also a coherent emitter \cite{de2025sporadic}. Therefore more constraints should be discussed. 

\textbf{Neutron Star ?}
We have demonstrated that DART J1832-0911 is likely a young neutron star that has been spun down due to interaction with an SNR fallback disk. Here, we further discuss this source under the assumption that it may be a spin-down powered neutron star. In a spin-down powered scenario, the magnetic field strength can be estimated using $B\sim 6.4\times 10^{19}$~G$\sqrt{P\dot{P}}$, and the spin-down luminosity is $\dot{E}=4\pi ^{2}I\dot{P}P^{-3}$~erg s$^{-1}$ where $I=10^{45}I_{45}~g~cm^{2}$ is the moment of inertia \cite{ostriker1969nature,contopoulos2006revised, contopoulos2014new}. Taking into account the estimated age of DART J1832-0911, which is constrained by the evolution timescale of the associated SNR to be between  $3\times 10^{4}$ and $2\times 10^{5}$ years, we can place limits on the magnetic field strength and spin-down luminosity. Specifically, the magnetic field strength limit is $B'>4\times 10^{16}$~G and the loss rate of rotational energy is $\dot{E}>4\times 10^{26}$~erg~s$^{-1}$.

We can compare the theoretical spin-down power to the emission power from the relation $L_{\textrm{radio}} \approx 1\times10^{19} (\frac{\textrm{sin}^{2}(\rho/2)}{\epsilon})(\frac{D}{\textrm{pc}}^{2}) (\frac{F}{\textrm{Jy}})(\frac{f}{\textrm{MHz}})$~erg/s, where $\rho$ is the angular radius of the emitting cone and $\epsilon$ is the pulse duty cycle\cite{szary2014radio}. We assume a peak flux of the emission at $F = 1$~Jy and the distance at $D=4.5$~kpc, taking a typical value\cite{szary2014radio} of $(\frac{\textrm{sin}^{2}(\rho/2)}{\epsilon})\approx10^{-2}$, then the radio luminosity then can be estimated as $L_{\textrm{radio}} \approx 8\times 10^{26}$~erg/s. 

We also discuss a spin-down mechanism caused by a supernova fallback disk. The braking process of a neutron star spun down by its surrounding supernova fallback disc is influenced by numerous aspects, including the initial mass $M_{\rm d0}$ and outer radius $R_{0}$ of the disc, the neutron star’s initial spin period $P_{0}$ and magnetic field strength $B_{0}$, with the instability of the fallback disc playing a critical role as well. The evolution of spin angular velocity $\Omega$ of the neutron star embedded in a fallback disk could be expressed as \cite{biryukov2021magnetic,yang2024instability}:
\begin{equation}
I\dot{\Omega} = N_{\rm acc}\cos \alpha + N_{\rm md} + N_{\rm dip}(1+\sin^2\chi)-\dot{I}\Omega,
\end{equation}
where $I$ is the moment of inertia of the NS, $\alpha$ and $\chi$ are the angles between the spin axis of the NS and rotation axis of the disk, and the magnetic inclination angle of the NS, respectively. The spin evolution of neutron stars assisted by fallback disks is mainly dominated by three torques: the accretion torque $N_{\rm acc}$, the torques caused by magnetic field-disk interaction $N_{\rm md}$ and by magnetic dipole radiation $N_{\rm dip}$. Among these, $N_{\rm md}$ plays a critical role in the formation of ultra-long-period pulsars. 

Moreover, it should be noted that radio emission can only be detected when neutron stars are in the radio ejector phase, which occurs when the inner radius of the disk exceeds the light cylinder radius or the disk becomes passive. Similar to accretion disks in dwarf novae and transient X-ray binaries, the fallback disks would be subject to the thermal ionization instability once the outermost regions of the disk cool sufficiently \cite{menou2001stability}. Under these conditions, the transport of material along the disk becomes inefficient, resulting in a passive disk and the cessation of its activity. The minimum temperature required for a disk to remain active is approximately 300 K \cite{ertan2009evolution}. Here we evolved one million systems, assuming that  the initial distributions of $M_{\rm d0}$, $R_{0}$, $B_{0}$ and $P_{0}$ follow as: $\log (M_{\rm d0}/M_{\odot})\in [-7,-1]$, $\log (R_{0}/R_{\rm s})\in [1,6]$, $\log B_{0}({\rm G})\in [13,16]$, and $\log P_{0}({\rm s})\in [-3,1]$. The simulated results of $P$ and $\dot{P}$ at $3\times 10^4$ yr and $2\times 10^5$ yr, corresponding to the age estimate of SNR G22.7-0.2, are shown as green and purple shaded areas in Fig.\ref{ppdot}. It is clear that these LPTs can be well covered by the simulated distributions of ejectors, which may support the hypothesis that such LPTs may originate from interactions between young neutron stars and fallback discs.

\begin{figure}
  \centering
  \includegraphics[width=1\linewidth]{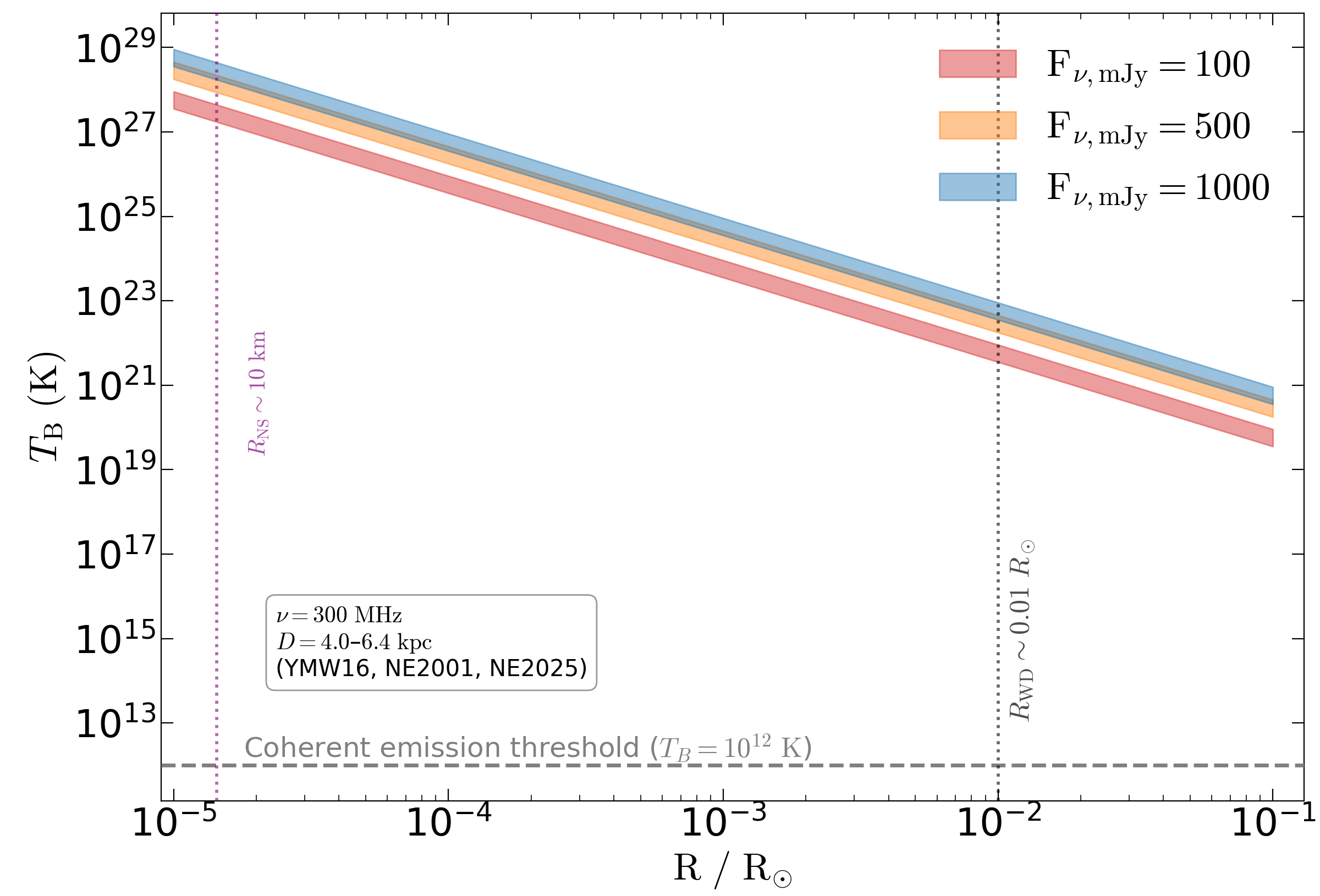}
 \caption{\textbf{The estimation of the radio brightness temperature against emitter radius.} The boundaries of each line correspond to the distance range from 4 to 6.4~kpc. The flux $\textrm{F}_{v,\textrm{mJy}}$ are estimated mean intensities.}
 \label{tb_r}
\end{figure}

\textbf{Magnetic White Dwarf ?}
The white dwarf model faces challenges in explaining DART J1832-0911s properties but cannot be entirely ruled out. The radio-limited brightness temperature $T_B$ provides constraints on the source quantities by examining the relationship between brightness temperatures and the size of the emitting region,  applicable to both thermal and coherent (non-thermal) emitters. Using the Rayleigh-Jeans approximation as a general definition of brightness temperature: 
\begin{equation}
T_{B} = \frac{c^2}{2\pi k}(\frac{D}{vR})^{2}F_{v}\approx 2\times 10^{19}v_{\textrm{MHz}}^{-2}F_{v,\textrm{mJy}}D_{\textrm{kpc}}^{2}\cdot (\frac{R}{R_{\odot}})^{-2},
\label{eq2}
\end{equation}
 where $R$ denotes the size of the emitting region (rather than the stellar radius). At $\nu = 300$~MHz and a distance range of $4 < D_{\rm kpc} < 6.4$~kpc (bracketing estimates from the YMW16, NE2001, and NE2025 electron-density models), 
the resulting $T_B$ as a function of emitting region size is 
depicted in Fig.~\ref{tb_r}.

 For any emitting region size comparable to a white dwarf radius (or much smaller), the inferred $T_B \sim 10^{19}$~K is characteristic of coherent emission, consistent with the broader LPT population as shown in Fig.\ref{phase_dig}. Since coherent emission is expected in both neutron star and magnetic white dwarf models \cite{marsh2016radio,pelisoli20235,de2025sporadic,hurley20242,dong2025chime,bloot2025strongly}, the $T_B$ alone 
does not discriminate between the two scenarios. However, the required brightness temperature still provides constraints. In confirmed WD-binary LPTs such as ILT~J1101+5521, the coherent radio emission arises from a large-scale interaction region between the two stars, with the emitting region on the order of the companion stellar radius ($\sim R_\odot$) \cite{de2025sporadic}, yielding $T_B \sim 10^{16}$~K. By contrast, the isolated nature of DART J1832-0911 implies that any WD model would require the coherent emission to originate from a compact region such as the WD magnetic pole ($R \lesssim 0.01~R_\odot$), pushing $T_B$ to $\gtrsim 10^{23}$~K. Such extreme brightness temperature implies a highly ordered and energetic particle beam, requiring a magnetic field strength well in excess of that observed in any known radio-emitting white dwarf 
($B \lesssim 10^9$~G for AR~Sco~\cite{marsh2016radio}), 
and more consistent with neutron star magnetospheric 
conditions.

 Another constraint comes from the energy budget and the characteristic cooling timescale.
Before the crystallization, the cooling timescale $\tau_{\textrm{cool}}$ of a hot WD is dominated by losing its thermal energy through black body radiation. Theoretically, the cooling time $\tau_{\textrm{cool}}$ is a function of the WD luminosity $L$, WD mass
$M_{\textrm{WD}}$, core atomic weight $A$. The approximate relation can be expressed as \cite{mestel1952theory,althaus2010evolutionary}
\begin{equation}
\tau_{\textrm{cool}} \approx \frac{10^{8}}{A}(\frac{M/M_{\odot}}{L/L_{\odot}})^{5/7}\qquad \textrm{years}
\end{equation}
where $M_{\odot}$ and $L_{\odot}$ are the solar mass and luminosity. The WD luminosity is $L=4\pi R_{\textrm{WD}}^{2}\cdot \sigma T_{\textrm{eff}}^{4}$. Assuming a typical WD with pure carbon
core, then $A=12$, the mass and radius are $M=0.6~M_{\odot}$ \cite{kepler2007white} and $R_{\textrm{WD}}=7,000$~km \cite{provencal1998testing}, respectively. 
We then obtained a cooling time $\tau_{\textrm{cool}}$ should be longer than $10^{8}$~years as $T_{\textrm{eff}}<17000$~K. 

Since WDs could be also spin-down powered dominated \cite{marsh2016radio}, we can estimate the characteristic age of a white dwarf candidate using the spin-down mechanism, which is given by:
\begin{equation}
\tau_{c}=\frac{P}{2\dot P}.
\end{equation} 
For such a radio-emitting WD , similar to the neutron star case \cite{geroyannis2000spin,buckley2017polarimetric}, we calculate the spin-down luminosity by adjusting for the WD's larger moment of inertia, typically $I'\approx10^{50}$~g $\textrm{cm}^{2}$ for a WD\cite{geroyannis2000spin}. 
Assuming that the radio emission of a WD is beamed like that of a pulsar, we then take the emission power $L_{\textrm{radio}}\approx8\times10^{26}$ erg/s as its rotational energy-loss rate. Then we derive the period derivative is $\dot P \sim 4\times 10^{-15}$ s/s. This gives a characteristic age of  $\tau_{c}\sim 10^{8}$ years, which is consistent with the lower limit cooling time scale based on the optical observation. However, such an analysis is parameter-dependent and serves only as a description of typical cases.

In conclusion, if a spin-down powered white dwarf is responsible for DART J1832-0911, it should be an extremely strongly magnetized, old and cold one.


\section*{Declarations}

\noindent\textbf{Conflict of interest.}
The authors declare that they have no conflict of interest.

\medskip
\noindent\textbf{Author contributions.}
D. Li, M. Yuan, and L. Wu conceived and designed the study.
J.-Y. Yan developed the detection pipeline/method and carried out the search.
X.-L. Lv coordinated the observations and collected the data.
All the other authors contributed to calibration and validation of the data, interpretation of the results, and preparation of the manuscript.

\medskip
\noindent\textbf{Data availability.}
The data that support the findings of this study are available from the corresponding author upon reasonable request.

\bibliographystyle{scibull}
\bibliography{SNbib_scibull}

\end{document}